\documentclass[fleqn,usenatbib]{mnras}

\usepackage{newtxtext,newtxmath}

\usepackage[T1]{fontenc}
\usepackage{ae,aecompl}
\usepackage{multirow}
\usepackage{subcaption}
\usepackage{enumerate}

\usepackage{graphicx}	
\usepackage{amsmath}	

\title[]{Synthesis of the first nitrogen-heterocycles in interstellar ice analogs containing methylamine (CH$_3$NH$_2$) exposed to UV radiation: Formation of trimethylentriamine (TMT, c-(-CH$_2$-NH)$_3$) and hexamethylentetramine (HMT, (CH$_2$)$_6$N$_4$).}

\author[H. Carrascosa et al.]{
H. Carrascosa,$^{1}$\thanks{E-mail: hcarrascosa@cab.inta-csic.es}
C. Gonz\'alez D\'iaz, $^{1}$
G. M. Mu\~noz Caro,$^{1}$\thanks{E-mail: munozcg@cab.inta-csic.es}
P. C. G\'omez,$^{2}$ 
M. L. Sanz, $^{3}$
\\
$^{1}$Centro de Astrobiolog\'{\i}a (CSIC-INTA), Ctra. de Ajalvir, km 4, Torrej\'on de Ardoz, 28850 Madrid, Spain\\
$^{2}$Dep. Qu\'imica F\'isica, Fac. Qu\'imica, Univ. Complutense, 28040 Madrid, Spain\\
$^{3}$Department of instrumental analysis and environmental chemistry, IQOG-CSIC, Juan de la Cierva 3, 28006 Madrid, Spain\\
}

\date{Accepted XXX. Received YYY; in original form ZZZ}

\pubyear{2021}

\begin{document}
\label{firstpage}
\pagerange{\pageref{firstpage}--\pageref{lastpage}}
\maketitle

\begin{abstract}
Hexamethylentetramine has drawn a lot of attention due to its potential to produce prebiotic species. This work aims to gain a better understanding in the chemical processes concerning methylamine under astrophysically relevant conditions. In particular, this work deeps into the formation of N-heterocycles in interstellar ice analogs exposed to UV radiation, which may lead to the formation of prebiotic species.\\
Experimental simulations of interstellar ice analogs were carried out in ISAC. ISAC is an ultra-high vacuum chamber equipped with a cryostat, where gas and vapour species are frozen forming ice samples. Infrared and ultraviolet spectroscopy were used to monitor the solid phase, and quadrupole mass spectrometry served to measure the composition of the gas phase.\\
The variety of species detected after UV irradiation of ices containing methylamine revealed the presence of 12 species which have been already detected in the ISM, being 4 of them typically classified as complex organic molecules: formamide (HCONH$_2$), methyl cyanide (CH$_3$CN), CH$_3$NH and CH$_3$CHNH. Warming up of the irradiated CH$_3$NH$_2$-bearing ice samples lead to the formation of trimethylentriamine (TMT), a N-heterocycle precursor of HMT, and the subsequent synthesis of HMT at temperatures above 230 K.\\
\end{abstract}


\section{Introduction}
\label{sect.introduction}



Molecules with six or more atoms including at least one C atom are known as complex organic molecules (COMs) \citep{Herbst2009ARAA..47..427H}. COMs have been detected in dense interstellar clouds and circumstellar regions. Deep inside dense clouds, the interstellar UV field cannot penetrate and dust temperature decreases to around 10 K. At these low temperatures, H$_2$O and other molecules including CO, CO$_2$, CH$_3$OH, CH$_4$, or NH$_3$ accrete onto dust grains forming ice mantles. These ice mantles are exposed to secondary-UV photons generated by the interaction between cosmic rays and hydrogen molecules, as well as direct cosmic ray impact, producing radicals and ions which can lead to the formation of new species. Some intermediate species can react at low temperature, while others remain in the ice until the temperature increases and thermal energy is enough to overcome the activacion barriers to form new species. Up to now, more than 200 different molecules have been detected in the gas phase toward interstellar and circumstellar environments, including around 50 COMs. An important fraction of these species is thought to be produced in ice mantles.\\

Among them, methylamine (CH$_3$NH$_2$) has drawn considerable attention due to its potential to produce prebiotic species. It was first detected in Sagitarius B2 and Orion A molecular clouds with column densities ranging from 1$\times$10$^{15}$ cm$^{-2}$ to 4$\times$10$^{15}$ cm$^{-2}$ \citep{Kaifu1974ApJ...191L.135K, Fourikis1974ApJ...191L.139F, Belloche2013AA...559A..47B}. \cite{Goesmann2015Sci...349b0689G} and \cite{Altwegg2017MNRAS.469S.130A} reported CH$_3$NH$_2$ abundances from 0.6\% to 1.2\% relative to H$_2$O on comet 67P/Churyumov-Gerasimenko. Finally, CH$_3$NH$_2$ has been also detected in hot cores. \cite{Ohishi201910.1093/pasj/psz068} reported the presence of CH$_3$NH$_2$ in G10.47+0.03 while \cite{Bogelund2019AA...624A..82B} observed column densities between 3.0$\times$10$^{15}$ and 2.7$\times$10$^{17}$ in three different hot cores from the high-mass star forming region NGC 6334I.\\

Different mechanisms have been proposed for methylamine formation under astrophysical conditions. In the gas phase, \cite{Herbst1985} modelled the possible association between CH$_3^+$ ions and NH$_3$ molecules to produce CH$_3$NH$_2$, concluding that it was an efficient route to produce methylamine under interstellar conditions, despite the existence of exothermic reaction channels. \cite{GARDNER1980353} and \cite{Ogura1989} reported the formation of methylamine from photolysis of gas mixtures containing CH$_4$ and NH$_3$. In their experiments, UV-irradiation produced CH$_3\cdot$ and NH$_2\cdot$ radicals able to recombine to produce CH$_3$NH$_2$. This pathway was also explored in ice samples by \cite{Kim2011ApJ...729...68K} and \cite{Forstel2017ApJ...845...83F}, who carried out electron bombardment and photon irradiation of CH$_4$:NH$_3$ ice mixtures, respectively, reporting the formation of methylamine in both works. An alternative route is the successive hydrogenation of gas phase HCN molecules, analogous to the formation of H$_2$CO and CH$_3$OH from CO molecules, as pointed out by \cite{Dickens1997ApJ...479..307D} and \cite{Theule2011AA...534A..64T}.\\

Several experiments reveal the potential of CH$_3$NH$_2$ to form prebiotic molecules. \cite{Holtom2005ApJ...626..940H} prepared CH$_3$NH$_2$:CO$_2$ binary ice mixtures which resulted in glycine (NH$_2$CH$_2$COOH) formation under electron bombardment. \cite{Bossa2009AA...506..601B} reproduced the same ice mixture in a water dominated ice. They found evidence on the formation of methylammonium methylcarbamate, [CH$_3$NH$_3^+$][CH$_3$NHCOO$^-$], a glycine salt precursor in astrophysical environments dominated by thermal and UV processing. In the same year, \cite{Lee2009ApJ...697..428L} reported the formation of several COMs from UV irradiation of CH$_3$NH$_2$:CO$_2$ mixtures on top of a H$_2$O ice.\\

The main motivation to study the UV-photoprocessing of methylamine-bearing ices in this work is to gain a better understanding on the formation of N-heterocycles in interstellar ice analogs. In particular, we targeted the synthesis of hexamethylentetramine (HMT), a molecule commonly identified among the refractory products in ice irradiation experiments, and its precursor, trimethylentriamine (TMT).\\

Hexamethylentetraamine (C$_6$N$_3$H$_{12}$, HMT) is a molecule of prime interest in astrochemistry. Firstly, it is a stable molecule efficiently made by UV or ion processing of interstellar ice analogs, which remains in the generated organic residue at room temperature \citep{Briggs1992OLEB...22..287B, Cottin2001ApJ...561L.139C}. Second, it may catalyze organic reactions leading to the formation of other COMs \citep{Vinogradoff2012}. Third, its acid hydrolysis leads to the formation of several aminoacids \citep{Fox1970Sci...170..984F}. HMT has been recently detected in Murchison (846$\pm$37 ppb), Murray (29$\pm$9 ppb) and Tagish Lake (671$\pm$9 ppb) carbonaceous chondrites, as well as the methyl, amino, hidroxy and hidroxymethyl derivatives of HMT \citep[see][]{Oba2020_0c90d3eeb3784052b68673613062d950}. The synthesis of HMT derivatives in ice irradiation experiments was previously reported \citep{Guille2004AA...413..209M, Sandford2020ChRev.120.4616S, Materese2020AsBio..20..601M,Urso2020AA...644A.115U}.\\

\begin{figure*}
  \centering 
  \includegraphics[width=1\textwidth]{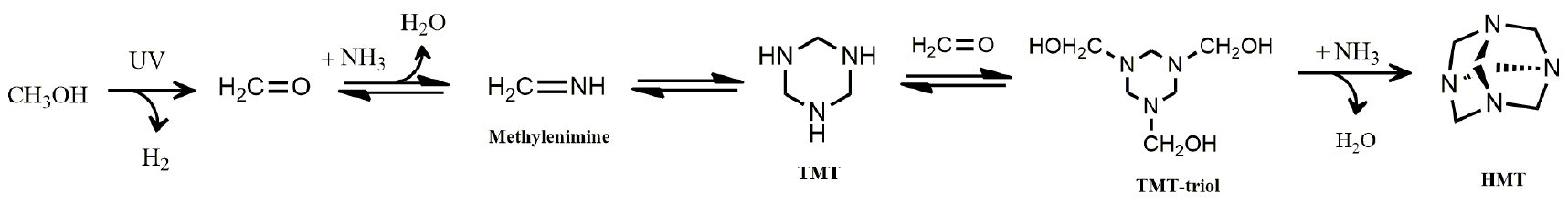}
  \caption{Synthetic pathway proposed for HMT synthesis.}
  \label{Fig.sintesis_HMT} 
\end{figure*}

HMT can be produced from condensation reaction between ammonia and formaldehyde in aqueous solution \cite[Fig. \ref{Fig.sintesis_HMT}]{Meissner1954doi:10.1021/ie50532a035}. This synthetic pathway has been typically applied to ice chemistry \cite[and references therein]{Bernstein1995, Guille2003AA...412..121M, Vinogradoff2013}. Basically, the proposed synthetic route is as follows. CH$_3$OH is easily dehydrogenated by UV radiation to produce formaldehyde, or formaldehyde is formed, albeit less efficiently, from hydrogenation of CO in the ice \citep{Briggs1992OLEB...22..287B, Guille2003AA...412..121M, Guille2004AA...413..209M}. The reaction between a carbonyl group and a primary amine typically leads to the formation of imine groups (R=N-R'), releasing a water molecule. Thus, from formaldehyde and ammonia, methylenimine (CH$_2$NH) is obtained. This highly reactive methylenimine polymerises, leading to a 6-member stable ring, known as trimethylentriazine (TMT). Methylen groups are electron-deficient units which incorporate hydroxymethyl radicals to form 1,3,5-trihydroxymethyltrimethylentriazine (TMT-triol). The latter leads to HMT in presence of ammonia. The last step, however, only takes place at high temperatures, when ammonia is already thermally desorbed to the gas phase in the ice irradiation and warm-up experiments. Indeed, \cite{Guille2003AA...412..121M} observed the formation of HMT at room temperature monitoring the growth of its main absorption bands in the IR spectra at 1007 and 1234 cm$^{-1}$. These authors proposed the presence of carboxylate ammonium salts, which would remain in the ice at high temperature, able to provide the amino groups.\\

\cite{Vinogradoff2012, Vinogradoff2013} and references therein deepened in the formation mechanism of HMT. Those works reported measurements of complex ice mixtures, confirming that formaldehyde reacts with ammonia to produce methylenimine (CH$_2$NH). However, the stability of organic rings favours the cyclation of the 6-member species CH$_2$NH-CH$_2$NH-CH$_2$NH, leading to the production of TMT \citep{Vinogradoff2012}. TMT is then converted into HMT near room temperature.\\

Within this work, we explore an alternative mechanism in the production of HMT. We have focused on the primary steps and the intermediate species. We have calculated the theoretical IR spectrum of TMT since we found no spectrum for this species in the literature. The validity of this calculation was checked comparing the calculated HMT spectrum to the one measured in the laboratory. \cite{Gardner1982} reported the preferential formation of CH$_2$NH from UV irradiation of CH$_3$NH$_2$ molecules, supported by the efficient formation of H$_2$ molecules. By using methylamine, we have been able to study in more depth the chemical pathways of C-N bearing molecules in different environments. Noble gas matrix isolation experiments were also carried out to isolate CH$_2$NH molecules, thus allowing the study of its spectroscopic features and radical formation. This will be explained in more detail in the paragraphs below.\\

In addition to the primary processes leading to HMT, we also studied the formation of other photoproducts arising upon UV irradiation of methylamine. When H$_2$O is added to CH$_3$NH$_2$ in the ice, in a likely more realistic astrophysical scenario, oxygenated species are formed. Some of them, such as HCN, HNCO or HCONH$_2$ were found in comet 67P/Churyumov-Gerasimenko surface during the Rossetta mission \citep{Goesmann2015Sci...349b0689G}, and confirmed by \cite{Altwegg2017MNRAS.469S.130A} using DFMS spectra from the orbiter. Finally, we produced a well-studied residue by irradiation of H$_2$O:CH$_3$OH:NH$_3$ ice mixture for comparison with the methylamine ice experiments \citep{Bernstein1995, Guille2003AA...412..121M, Vinogradoff2013}\\

The present paper is structured as follows. Sect. \ref{sect.experimental_setup} explains the experimental procedure used during the experimental simulations, as well as the calculated IR spectra. Sect. \ref{Results} is divided into four subsections. Sect \ref{sect.metilamina_environment} presents the different features of CH$_3$NH$_2$ in the three environments studied in this work. Sect. \ref{Sect.H2_diffussion} explores the different rate of H$_2$ subtraction during UV irradiation of the different ice matrices. Sect. \ref{sect.fotoproductos} focuses on the formation of photoproducts at 8 K in the different ice mixtures. Sect. \ref{sect.TMT} goes deeper into the formation of large species, such as TMT, which only takes place at higher temperatures. Finally, Sect. \ref{sect.conclusions} and Sect. \ref{sect.astrophysical_implications} summarizes the main results and present the astrophysical implications.\\

\section{Experimental}
\label{sect.experimental_setup}

\subsection{ISAC experimental simulations}
Experiments were carried out in the Interstellar Astrochemistry Chamber (ISAC), fully described in \cite{Guille2010}. ISAC is an Ultra-High Vacuum (UHV) chamber with a base pressure of 4$\times$10$^{-11}$ mbar equipped with a closed-cycle helium cryostat able to reach temperatures of 8 K in a MgF$_2$/KBr substrate used for ice deposition.\\

For the experiments, highly distilled MilliQ water was obtained from a Millipore water distribution system IQ-7000. For matrix experiments, Xe was acquired from Praxair with a 99.999\% purity. Methylamine was purchased from Merck group diluted in water (40\% in water solution). The higher vapour pressure of methylamine compared to the solvent allowed us to prepare relatively pure samples, as confirmed by infrared spectroscopy, although traces of water ($\leq5\%$) are assumed to be present in "pure" methylamine ices. Nevertheless, the low deposition rate in Xe isolated methylamine ice samples determined the absence of any water signals in the IR spectrum. For H$_2$O:CH$_3$OH:NH$_3$ ice mixtures, CH$_3$OH was purchased from Panreac Quimica S.A. (99.9\% purity) and NH$_3$ from Air Liquide Espa\~na S.A. ($\geq$99.96\% purity).\\

Gases and vapours were introduced in the main chamber through a capillary tube of 1 mm internal diameter at normal incidence angle with respect to the MgF$_2$ substrate at pressures ranging from 2$\times$10$^{-7}$ to 1$\times$10$^{-6}$ for the different experiments. To ensure the purity of the ice mixtures, a quadrupole mass spectrometer (QMS, Pfeiffer Vacuum, Prisma QMS 200) was placed in the injection system, thus avoiding any contamination and providing a method to achieve the desired ratio between the components. The gas phase was monitored continuously in the main chamber using another QMS (Pfeiffer Vacuum, Prisma QMS 200) equipped with a Channeltron detector. For matrix isolation experiments, Xe pressure (monitored by $\frac{m}{z}$ = 131) was adjusted to measure an ion current at least two orders of magnitude larger than the one measured for the most intense fragment from methylamine ($\frac{m}{z}$ = 30). In addition, Xe has three natural isotopes with similar abundances, thus, the real ratio Xe:CH$_3$NH$_2$ turned out to be around 10.000:1, providing a good isolation of CH$_3$NH$_2$ molecules. This is supported by the dissapearance of the N-H stretching vibrations in the IR spectrum, as explained in Sect. \ref{sect.metilamina_environment}.\\

Fourier-transform infrared spectroscopy (FTIR) transmittance spectra were recorded using a Bruker VERTEX 70 with a resolution of 1-2 cm$^{-1}$, equipped with a deuterated triglycine sulfate (DGTS) detector to monitor the solid phase. IR spectra of the ice samples were performed before and after deposition of the ice, after each irradiation interval, and during the warming up. From the infrared spectra, the column density was estimated following eq. \ref{Eq.Band.strength}, where $A$ is the band strength in cm molecule$^{-1}$, $\tau_{\nu}$ is the optical depth, and $d\nu$ is the wavenumber differential in cm$^{-1}$. For methylamine, the adopted band strength is the one reported by \cite{Holtom2005ApJ...626..940H}, 4.3$\times$10$^{-18}$ cm molecule$^{-1}$ for the 1613 cm$^{-1}$ IR band. For H$_2$O, a value of 2.0$\times$10$^{-16}$ cm molecule$^{-1}$ for its 3259 cm$^{-1}$ IR band was used \citep{Hagen1981CP.....56..367H}. For methanol, 1.8$\times$10$^{-17}$ cm molecule$^{-1}$ for its 1025 cm$^{-1}$ IR band \citep{d'Hendecourt1986AA...158..119D}. Finally, for NH$_3$, a value of 1.7$\times$10$^{-17}$ cm molecule$^{-1}$ for its 1070 cm$^{-1}$ IR band was adopted \citep{sandford1993ApJ...417..815S}.\\

\begin{equation}
\centering
N = \frac{1}{A} \int_{band}{\tau _{v}\; dv.}
\label{Eq.Band.strength}
\end{equation}\\ 

For optically thick ices, the ratio between the components was estimated from the ion current measured by the QMS during the deposition of the ice samples, which showed a good agreement with the values calculated from IR spectroscopy.\\

Ice samples were irradiated with a F-type microwave discharged hydrogen lamp (MDHL) from Opthos instruments. The light emitted by the MDHL enters the ISAC chamber through a MgF$_2$ window, and radiation is guided to the deposition substrate through a quartz tube inside the chamber. The UV-flux was measured at the end of the quartz tube with a calibrated Ni mesh to know the flux at the sample position \citep[see][]{Cristobal2019}. Vacuum ultraviolet spectroscopy was carried out using a McPherson 0.2 m focal length UV monochromator (model 234/302) placed in front of the MDHL. The UV-absorption cross section of the ice samples was obtained applying eq. \ref{Eq.VUV_asborption}, where $I_t(\lambda)$ is the transmitted intensity, $I_0(\lambda)$ is the incident intensity, $\sigma(\lambda)$ is the UV-absorption cross section of the ice, and $N$ is the column density derived from the IR spectra, following \cite{Gus2014}.\\

\begin{equation}
\centering
I_t(\lambda) = I_0(\lambda) \times e^{-\sigma(\lambda) N}
\label{Eq.VUV_asborption}
\end{equation}\\ 

After ice irradiation, temperature programmed desorption (TPD) experiments were performed. A LakeShore 331 temperature controller connected to a silicon diode sensor with a sensibility better than 0.1 K was used to warm up the ice samples at 0.3 K/min up to 300 K. QMS data were recorded continuously and FTIR spectra were measured every 5 K during warm-up.\\

For H$_2$O:CH$_3$OH:NH$_3$ experiments, H$_2$O and CH$_3$OH were introduced through the main gas line in ISAC, while NH$_3$ was introduced through a second capillary tube, using a secondary gas line specifically designed for corrosive species.\\

\subsection{IR spectra simulations}
The calculation of the vibrational spectra for all the species collected in Fig. \ref{Fig.simulaciones} has been done at the harmonic level using the well-tested Density Functional Theory hybrid method B3LYP. Different basis sets were used in exploratory calculations and finally the quadruple-z quality correlation consisten aug-cc-pVQZ, which includes both polarization and diffuse functions, was used for production purposes. Frequencies presented are not scaled and intensities are referred to the highest feature in each spectrum. All the calculations were carried out by means of Gaussian16 package \citep{gaussian16}.
\section{Results and discussion}
\label{Results}

\subsection{Theoretical IR spectra}

\begin{figure}
  \centering 
  \includegraphics[width=0.5\textwidth]{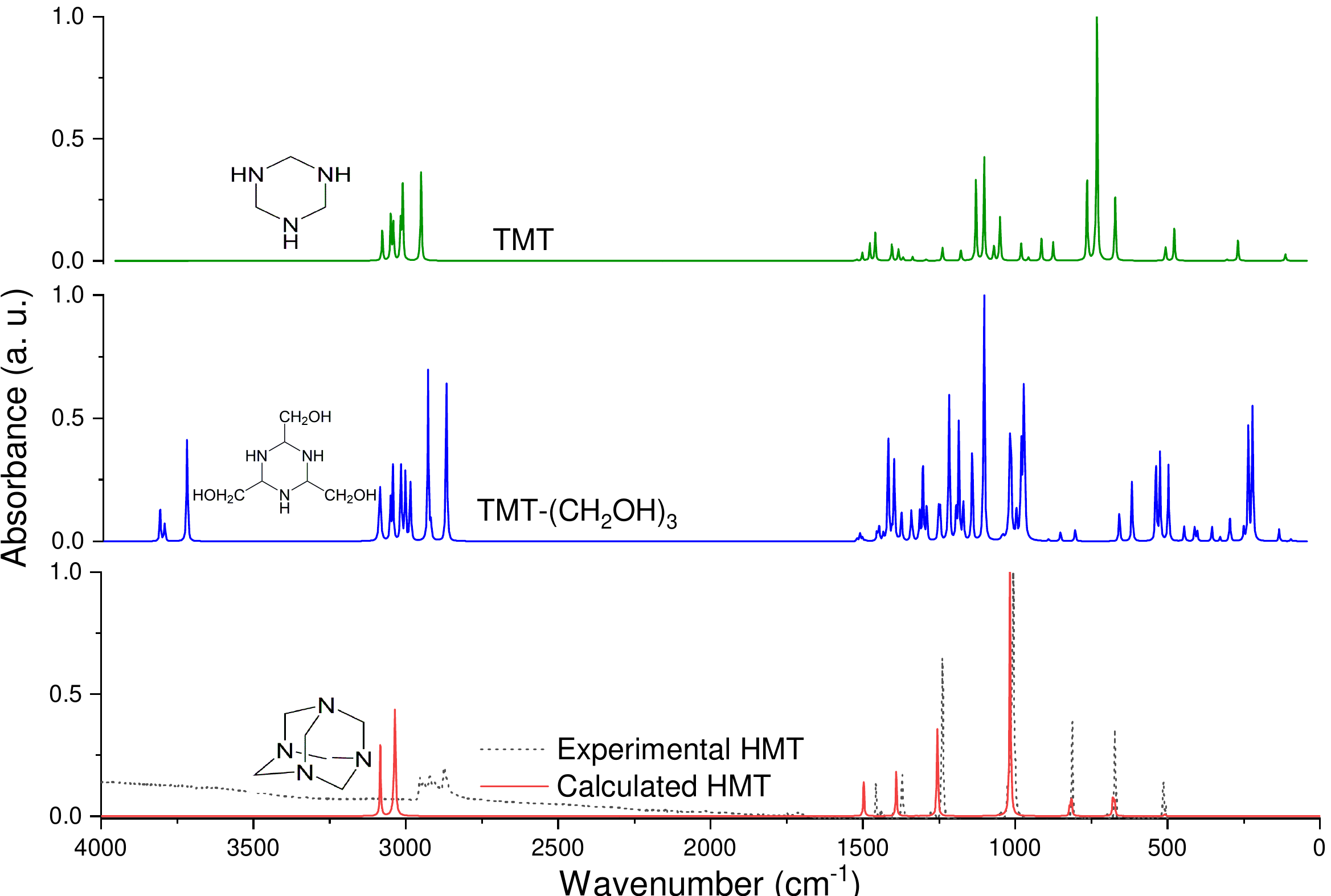}
  \caption{Calculated IR spectra of three species related to HMT synthesis: TMT, TMT-(CH$_2$OH)$_3$ and HMT itself. Dash black line shows the experimental IR spectra of HMT for comparison. Absorbance was normalized to 1 in all cases.}
  \label{Fig.simulaciones}
\end{figure}

Fig. \ref{Fig.simulaciones} shows the calculated IR spectra of three species related to HMT synthesis: TMT, TMT-(CH$_2$OH)$_3$ and HMT. The infrared spectra of TMT and TMT-(CH$_2$OH)$_3$ have, to our knowledge, not been reported before. The well known IR spectrum of HMT was also calculated to illustrate the reliability of the synthetic spectra. There is a good agreement between our measured IR spectrum of HMT and the calculated one, only the shift of the CH$_x$ stretching modes near 3000 cm$^{-1}$, also present in the calculated IR spectrum of HMT derivatives \citep{Bera2019ApJ...884...64B}, is worth noting.\\

\subsection{Effect of the ice environment over CH$_3$NH$_2$ ice.}
\label{sect.metilamina_environment}

\begin{figure}
  \centering 
  \includegraphics[width=0.5\textwidth]{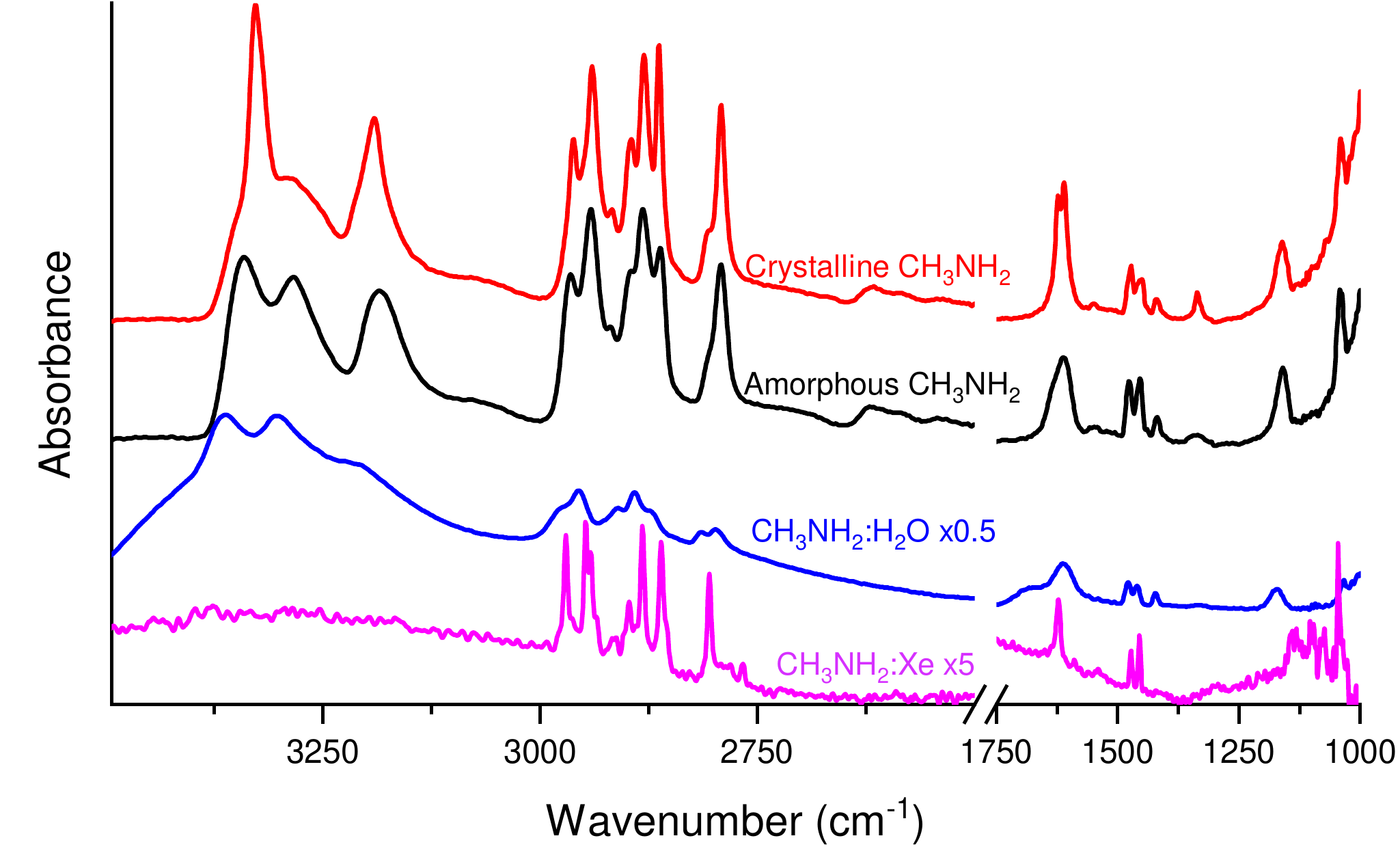}
  \caption{IR spectra recorded after deposition of different ice samples. CH$_3$NH$_2$:Xe (pink, Exp. \textbf{5}), CH$_3$NH$_2$:H$_2$O (blue, Exp. \textbf{9}), and pure CH$_3$NH$_2$ at 8 K (black, Exp. \textbf{1}) and 110 K (red, after crystallization process in Exp. \textbf{1}). Band positions agree with those reported for pure CH$_3$NH$_2$ ice \citep{Durig1968JChPh..49.2106D, Zhu2019PCCP...21.1952Z}.}
  \label{Fig.IR_metilamina_no_irradiada}
\end{figure} 

Table \ref{Table.experiments} collects the different ice sample compositions, ice column density, and irradiation dose of the experiments presented in this work. Fig. \ref{Fig.IR_metilamina_no_irradiada} shows the methylamine IR features observed after deposition of CH$_3$NH$_2$ molecules in different environments (Exp. \textbf{1}, \textbf{5} and \textbf{6}). Pure methylamine ice exhibits features (shown in Table \ref{Table.IR_metilamina_inicial}) which are in good agreement with previous works \citep[and references therein]{Durig1968JChPh..49.2106D, Zhu2019PCCP...21.1952Z}. When CH$_3$NH$_2$ is diluted in an inert matrix, Xe atoms in this work, intermolecular forces are reduced, and bands become narrower. N-H stretching vibrations (3500-3100 cm$^{-1}$) seem to disappear in the Xe matrix, indicative of a large decrease of its IR band strength when molecules are isolated and H bonds are inhibited \citep{Durig2001_CH3NH2_Kr_Xe}. CH$_3$NH$_2$ in a CH$_3$NH$_2$:H$_2$O (1:1) ice mixture (Exp. \textbf{6}) shows the same IR features observed for pure CH$_3$NH$_2$ ice, slightly red-shifted by the different intermolecular forces present in this ice, and widened, as a consequence of the interaction of CH$_3$NH$_2$ with H$_2$O neighbouring molecules. N-H stretching bands, however, are not weakened in a H$_2$O dominated ice. O-H groups in water ice can interact strongly with N-H groups, as both N and O atoms are able to establish hydrogen bonds.\\

\begin{table} 
    \caption[]{Experiments performed in this work}
    \label{Table.experiments}
    \resizebox{8.5cm}{!}{
    \begin{tabular}{ccccc}
Exp.         & Ice sample        & N$_0$ (CH$_3$NH$_2$)         & \multicolumn{2}{c}{Photon dose}\\
\hline
            &                   &                               &Absolute       &per molecule\\
            &                   & (cm$^{-2}$)                   & (cm$^{-2}$)   & (cm$^{-2}$)\\
\noalign{\smallskip}
\hline
\noalign{\smallskip}
\textbf{1}&CH$_3$NH$_2$         & 2.4$\times$10$^{17}$            & 0                        &0\\
\noalign{\smallskip}
\textbf{2}&CH$_3$NH$_2$         & 3.2$\times$10$^{17}$         & 1.0$\times$10$^{18}$     &3.2\\
\noalign{\smallskip}
\textbf{3}&CH$_3$NH$_2$         & 6.1$\times$10$^{17}$         & 1.4$\times$10$^{18}$     &2.5\\
\noalign{\smallskip}
\textbf{4}&CH$_3$NH$_2$         & 1.2$\times$10$^{18}$         & 1.5$\times$10$^{18}$     &1.1\\
\noalign{\smallskip}
\noalign{\smallskip}
\textbf{5}&CH$_3$NH$_2$:Xe**   & 2.2$\times$10$^{16}$         & 1.0$\times$10$^{18}$     &-\\
\noalign{\smallskip}
\textbf{6}&CH$_3$NH$_2$:H$_2$O (1:1)  & 3.9$\times$10$^{17}$         & 9.0$\times$10$^{17}$     &1.2\\
\noalign{\smallskip}
\textbf{7}&CH$_3$NH$_2$:H$_2$O* (2:1)  & 6.0$\times$10$^{17}$         & 2.7$\times$10$^{18}$     &1.5\\
\noalign{\smallskip}
\textbf{8}&CH$_3$NH$_2$         & 1.9$\times$10$^{19}$         & 3.6$\times$10$^{19}$     &1.9\\
\noalign{\smallskip}
\textbf{9}&CH$_3$NH$_2$:H$_2$O (1:2)    & 2.0$\times$10$^{18}$         & 1.5$\times$10$^{18}$     &0.25\\
\hline
\end{tabular}\\
}
\scriptsize{* Experiment \textbf{7} was irradiated during the warm-up, between 8 K and 110 K. This irradiation represents the 40\% of the total photon dose.\\
** Absorption of Xe atoms prevented us from the quantification of methylamine photon dose, see Fig. \ref{Fig.VUV_comparison}.}\\ 
\end{table}

\begin{table} 
\caption[]{IR features of CH$_3$NH$_2$ in the different ice mixtures. For pure CH$_3$NH$_2$ ice, data agree with \cite{Durig1968JChPh..49.2106D, Zhu2019PCCP...21.1952Z}. For CH$_3$NH$_2$:Xe and CH$_3$NH$_2$:H$_2$O, no reference was found in the literature.}
    \label{Table.IR_metilamina_inicial}
\resizebox{8.5cm}{!}{
\begin{tabular}{llll}
Vibration mode  & CH$_3$NH$_2$:Xe & CH$_3$NH$_2$ & CH$_3$NH$_2$:H$_2$O\\
\hline
\hline
\noalign{\smallskip}
NH$_2$ antisymmetric str.                    &-             & 3347             &3354  \\
\noalign{\smallskip}
NH$_2$ symmetric str.                        &-             & 3286             &3286  \\
\noalign{\smallskip}
H bonding                                    &-             &   3182           &3186  \\
\noalign{\smallskip}
CH$_3$ degenerate str.                       &2970,2942     &   2968, 2944     &2968, 2945  \\
\noalign{\smallskip}
CH$_3$ degenerate str.                       &2882, 2861    &   2884, 2864     &2919, 2878  \\
\noalign{\smallskip}
CH$_3$ symmetric str.                        &2805, 2766    &   2808, 2793     &2809, 2792  \\
\noalign{\smallskip}
NH$_2$ scissoring                            &1622          & 1613            &1612  \\
\noalign{\smallskip}
CH$_3$ degenerate bend                       &1472, 1455    &   1476, 1457    &1478, 1457  \\
\noalign{\smallskip}
NH$_2$ twist                                 &-             &   1418          &1421  \\
\noalign{\smallskip}
                                             & -            &  1337           & - \\
\noalign{\smallskip}
CH$_3$ rocking                               &-             &   1161          &1166  \\
\noalign{\smallskip}
C-N str.                                     &1045          &   1041          &1040 \\
\noalign{\smallskip}
NH$_2$ wagging                               & -            &  995            & -  \\
\noalign{\smallskip}
\noalign{\smallskip}
\hline
\end{tabular}\\
}
\textit{str. = stretching; bend. = bending}
\end{table}

\begin{figure}
  \centering
  \includegraphics[width=0.5\textwidth]{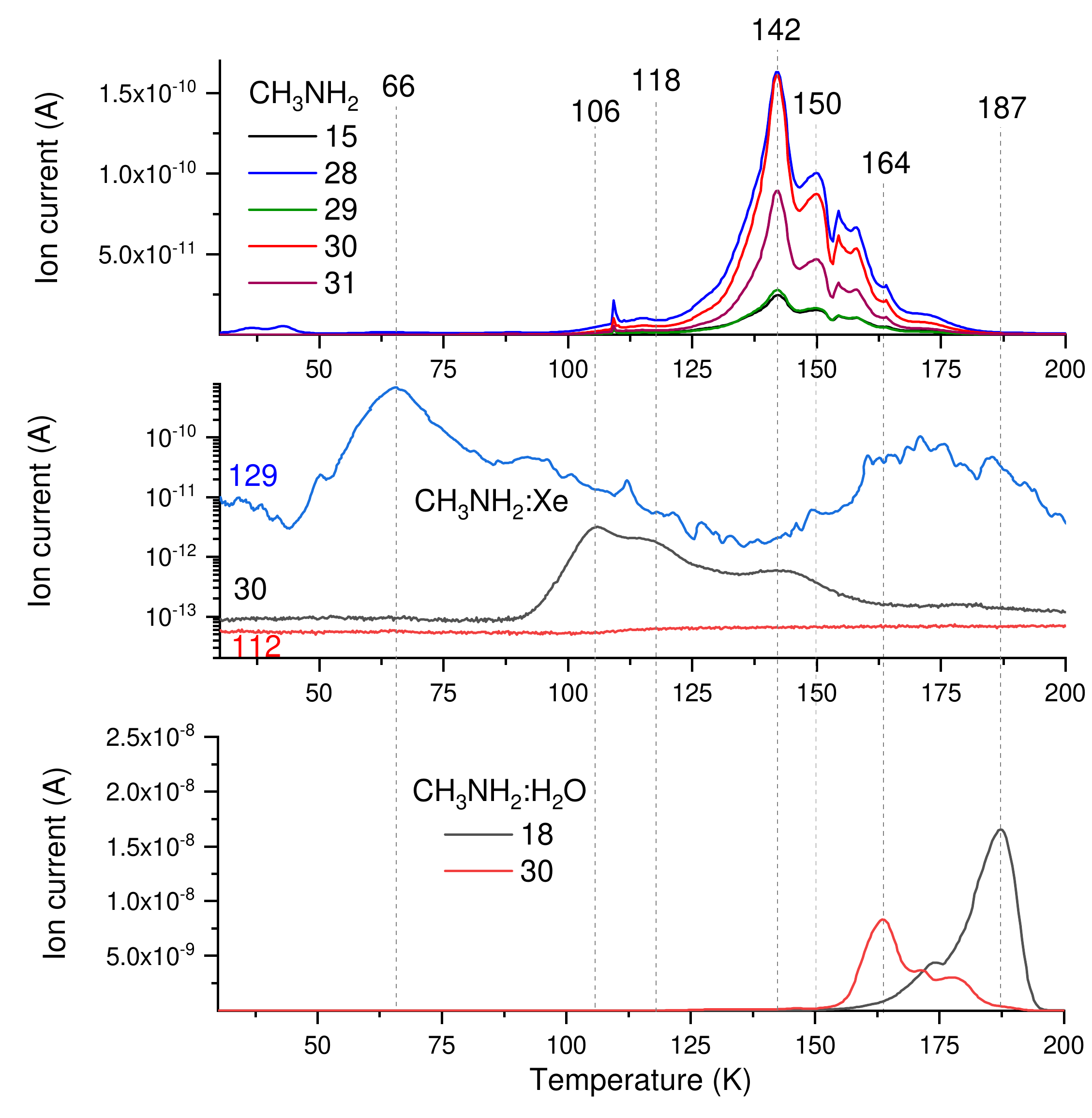}
  \caption{Top: thermal desorption of unirradiated CH$_3$NH$_2$ ice leading to ejection of these molecules to the gas phase. The most intense $\frac{m}{z}$ fragments of CH$_3$NH$_2$ are recorded. Middle: thermal desorption of Xe and CH$_3$NH$_2$ during warm-up of experiment \textbf{6}. No signal is expected for $\frac{m}{z}$ = 112, it is shown for comparison. Bottom: thermal desorption of CH$_3$NH$_2$ and H$_2$O during warming up of experiment \textbf{10}.}
  \label{Fig.QMS_metilamina_environment}
\end{figure} 

TPD experiments over a non-irradiated CH$_3$NH$_2$ ice revealed no major changes in the IR spectra up to 100 K. The onset of CH$_3$NH$_2$ crystallization occurs near 100 K, CH$_3$NH$_2$ acquires a crystalline structure (Fig. \ref{Fig.IR_metilamina_no_irradiada}), ejecting molecules with a low-binding energy during this process. Fig. \ref{Fig.QMS_metilamina_environment} shows the thermal desorption of CH$_3$NH$_2$ in the different environments studied in this work. Pure methylamine ice sublimates from 100 K, having its maximum desorption at 142 K. The desorbed molecules are responsible for the QMS peak starting at 90 K with a maximum at 106 K in the Xe isolated methylamine ice (middle panel of Fig. \ref{Fig.QMS_metilamina_environment}). When Xe atoms thermally desorb at 66 K, the remaining methylamine molecules do not strongly interact. Intermolecular forces are reduced, and thermal desorption takes place at lower temperature corresponding to the onset temperature of desorption in the pure CH$_3$NH$_2$ ice shown in the top panel of Fig. \ref{Fig.QMS_metilamina_environment}. Co-deposition with less-volatile H$_2$O molecules delays CH$_3$NH$_2$ thermal desorption to 164 K. Indeed, some methylamine molecules remain in the solid phase at larger temperatures, desorbing during the crystallization of H$_2$O molecules, around 175 K. From Fig. \ref{Fig.QMS_metilamina_environment}, it can be concluded that methylamine thermal desorption is strongly dependent on its environment, as it happens with the type of substrate when dealing with monolayer-thick ices \citep{Chaabouni2018AA...612A..47C}.\\

\subsection{UV irradiation of CH$_3$NH$_2$ ice samples. H$_2$ formation and diffusion. Saturation of species}
\label{Sect.H2_diffussion}

\begin{figure}
  \centering 
  \includegraphics[width=0.5\textwidth]{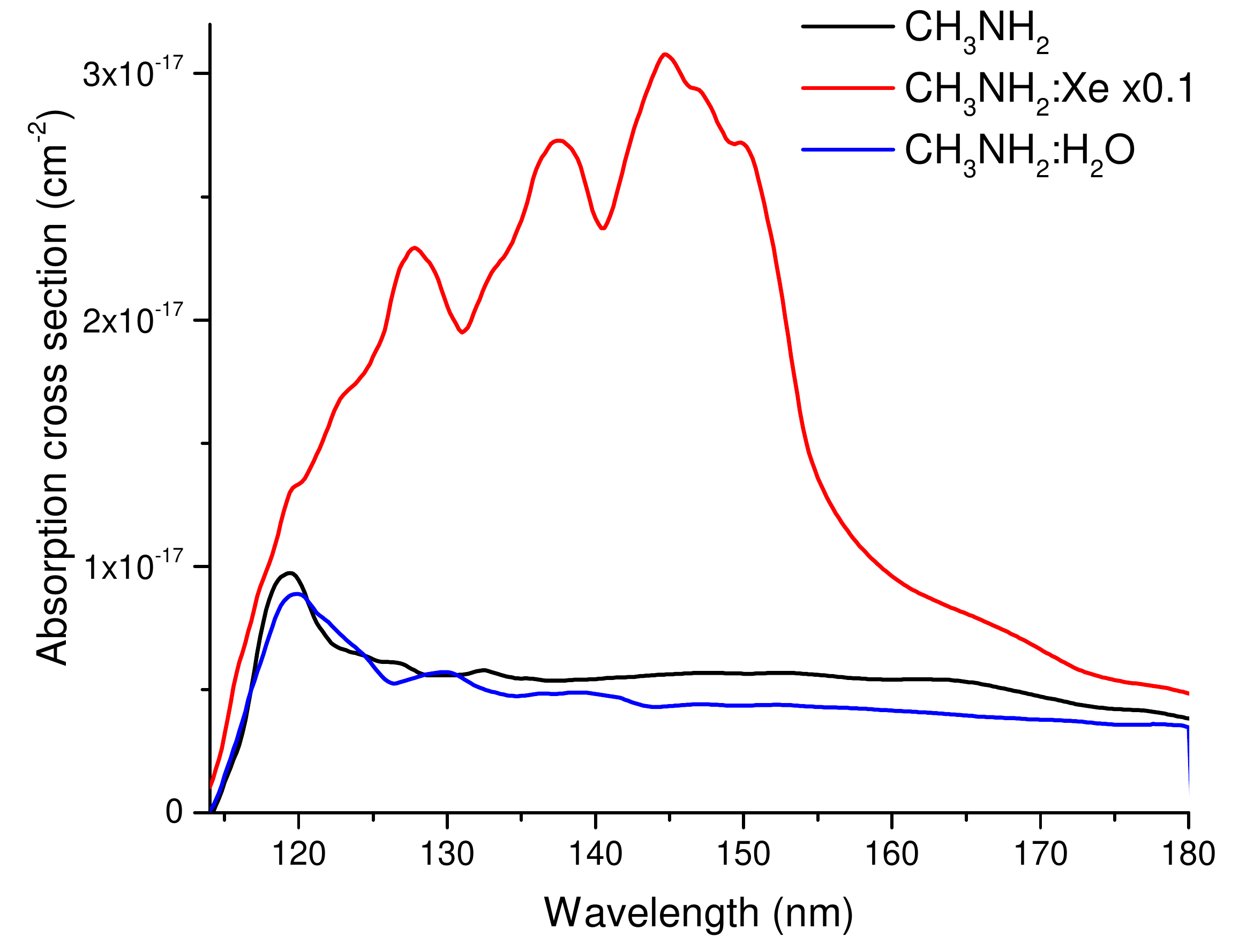}
  \caption{UV spectra of pure CH$_3$NH$_2$ (black trace), CH$_3$NH$_2$:Xe (red trace), and CH$_3$NH$_2$:H$_2$O (blue trace) ice samples. Note that Xe atoms dominate the UV absorption in the CH$_3$NH$_2$:Xe ice sample, see text.}
  \label{Fig.VUV_comparison}
\end{figure} 

As it can be seen in Table \ref{Table.experiments}, the column density of CH$_3$NH$_2$ in the pure ice samples was larger by a factor of $\geq$ 10 compared to CH$_3$NH$_2$:Xe ice sample (Exps. \textbf{2} and \textbf{6}). From the UV absorption cross section of methylamine presented in Fig. \ref{Fig.VUV_comparison}, and using eq. \ref{Eq.VUV_asborption}, a column density of 6.9$\times$10$^{17}$ cm$^{-2}$ corresponds to about 95\% UV absorption for pure methylamine ice, and 4.2$\times$10$^{17}$ cm$^{-2}$ for a 95\% UV absorption in the CH$_3$NH$_2$:Xe ice sample. Therefore, in our experiments, most of the CH$_3$NH$_2$ molecules in the ice were exposed to UV radiation. UV-thick ices were irradiated during the deposition process, their corresponding photon dose is shown in Table \ref{Table.experiments}. The large difference between CH$_3$NH$_2$ (both pure and in H$_2$O mixture) with the UV spectrum of CH$_3$NH$_2$:Xe ice mixture (see Fig. \ref{Fig.VUV_comparison}) is due to the UV absorption of Xe ice, in line with the Xe absorption presented by \cite{schnepp1960JChPh..33...49S}.\\

\begin{figure}
  \centering
  \includegraphics[width=0.5\textwidth]{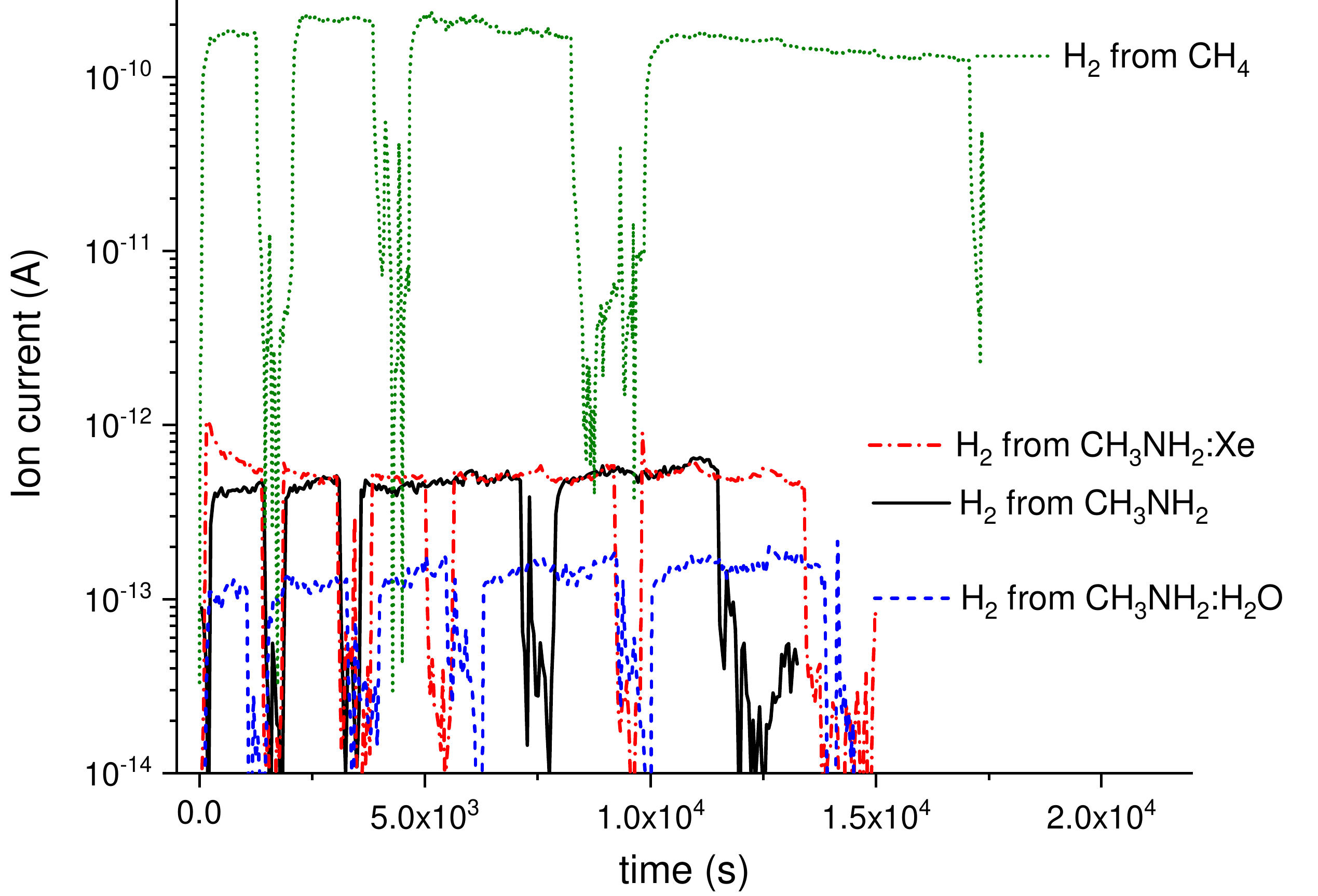}
  \caption{H$_2$ signal recorded during irradiation of CH$_3$NH$_2$ pure (Exp. \textbf{2}), CH$_3$NH$_2$:Xe (Exp. \textbf{5}), and CH$_3$NH$_2$:H$_2$O (Exp. \textbf{7}) ice samples. H$_2$ signal from irradiation of CH$_4$ ice is shown for comparison \citep{Carrascosa2020MNRAS.493..821C}. Fast changes between the larger and the lower H$_2$ signal are produced when the UV lamp is turned on and off along the irradiation period. Horizontal lines are drawn to guide the eye. Note that the y-axis is in logarithmic scale.}
  \label{Fig.comparison_H2_production}
\end{figure}

\begin{figure}
  \centering 
  \includegraphics[width=0.5\textwidth]{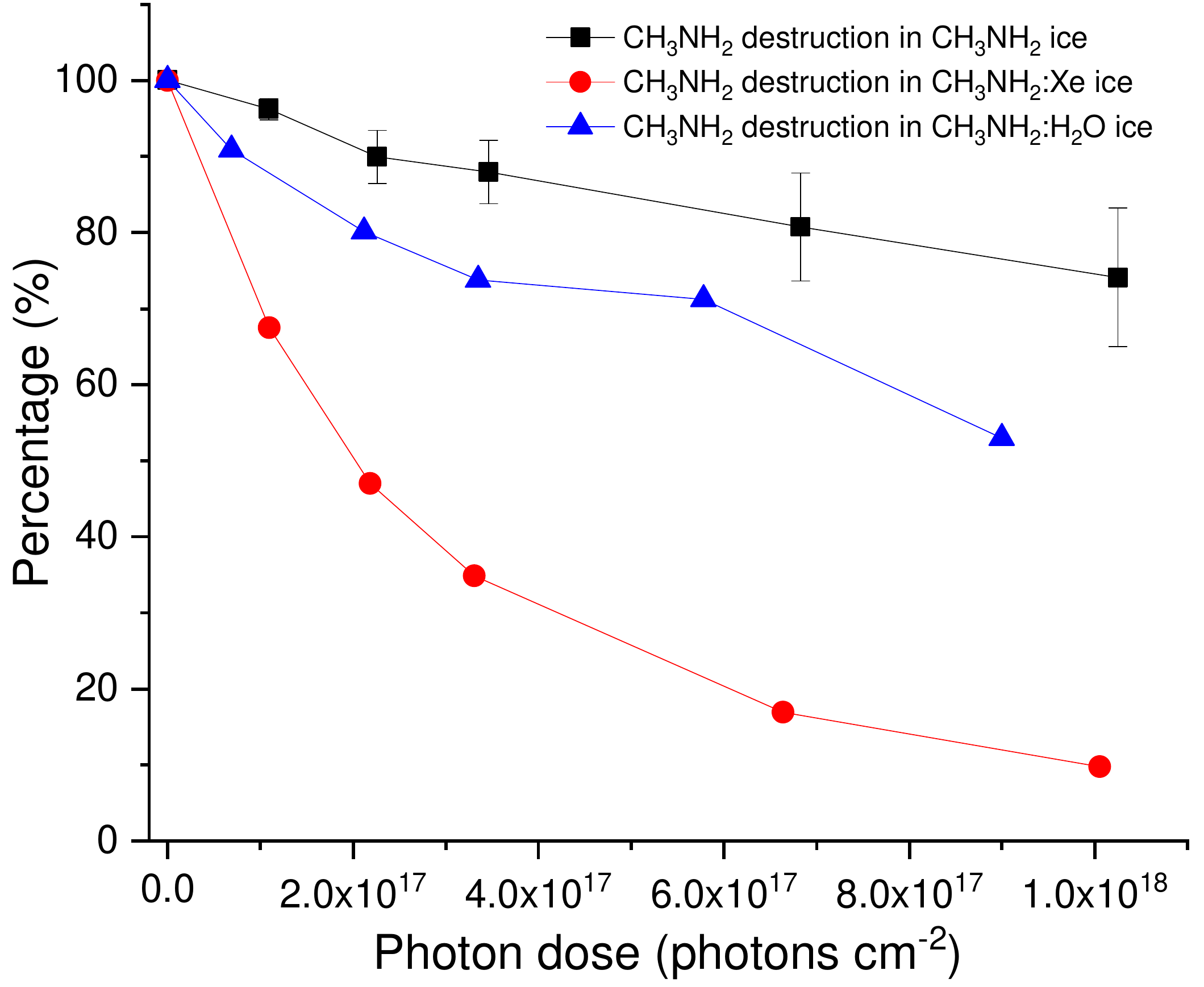}
  \caption{Decrease in the CH$_3$NH$_2$ column density after each irradiation interval for CH$_3$NH$_2$ (Exp. \textbf{2}), CH$_3$NH$_2$:Xe (Exp. \textbf{5}) and CH$_3$NH$_2$:H$_2$O (Exp. \textbf{7}).}
  \label{Fig.destruccion_metilamina}
\end{figure}

H$\cdot$ subtraction is the main process concerning UV irradiation of methylamine. The high mobility of hydrogen atoms in the ice bulk and surface allows the fast formation of H$_2$ molecules. Diffusion and eventual ejection of H$_2$ molecules enables its detection by QMS, where a larger signal is expected for the pure CH$_3$NH$_2$ ice when compared to CH$_3$NH$_2$:Xe ice mixture, due to its larger CH$_3$NH$_2$ column density. Nevertheless, Fig. \ref{Fig.comparison_H2_production} shows that ion current measured for $\frac{m}{z} = 2$, related to desorbing H$_2$ molecules is similar in both experiments. The formation of hydrogen bonds in the ice avoids hydrogen diffusion. Therefore, pure CH$_3$NH$_2$ ice shows a lower H$_2$ diffusion compared to matrix isolated CH$_3$NH$_2$. The addition of H$_2$O allows the formation of more hydrogen bonds, reducing H$_2$ diffusion even further. A lower H$_2$ diffusion makes H$\cdot$ atoms and H$_2$ molecules remain longer in the CH$_3$NH$_2$ and the CH$_3$NH$_2$:H$_2$O ice samples, enhancing chemical reactions with $\cdot$CH$_2$NH$_2$ and CH$_3$NH$\cdot$ radicals to form back CH$_3$NH$_2$. This recombination leads to a lower than expected H$_2$ signal in the QMS (Fig. \ref{Fig.comparison_H2_production}), as well as to a lower overall dissociation of CH$_3$NH$_2$ molecules during UV irradiation, shown in Fig. \ref{Fig.destruccion_metilamina}. In the CH$_3$NH$_2$:Xe ice samples, hydrogen diffusion is larger, and recombination is quenched.\\

\begin{figure}
\begin{subfigure}{.45\textwidth}
  \includegraphics[width=\linewidth]{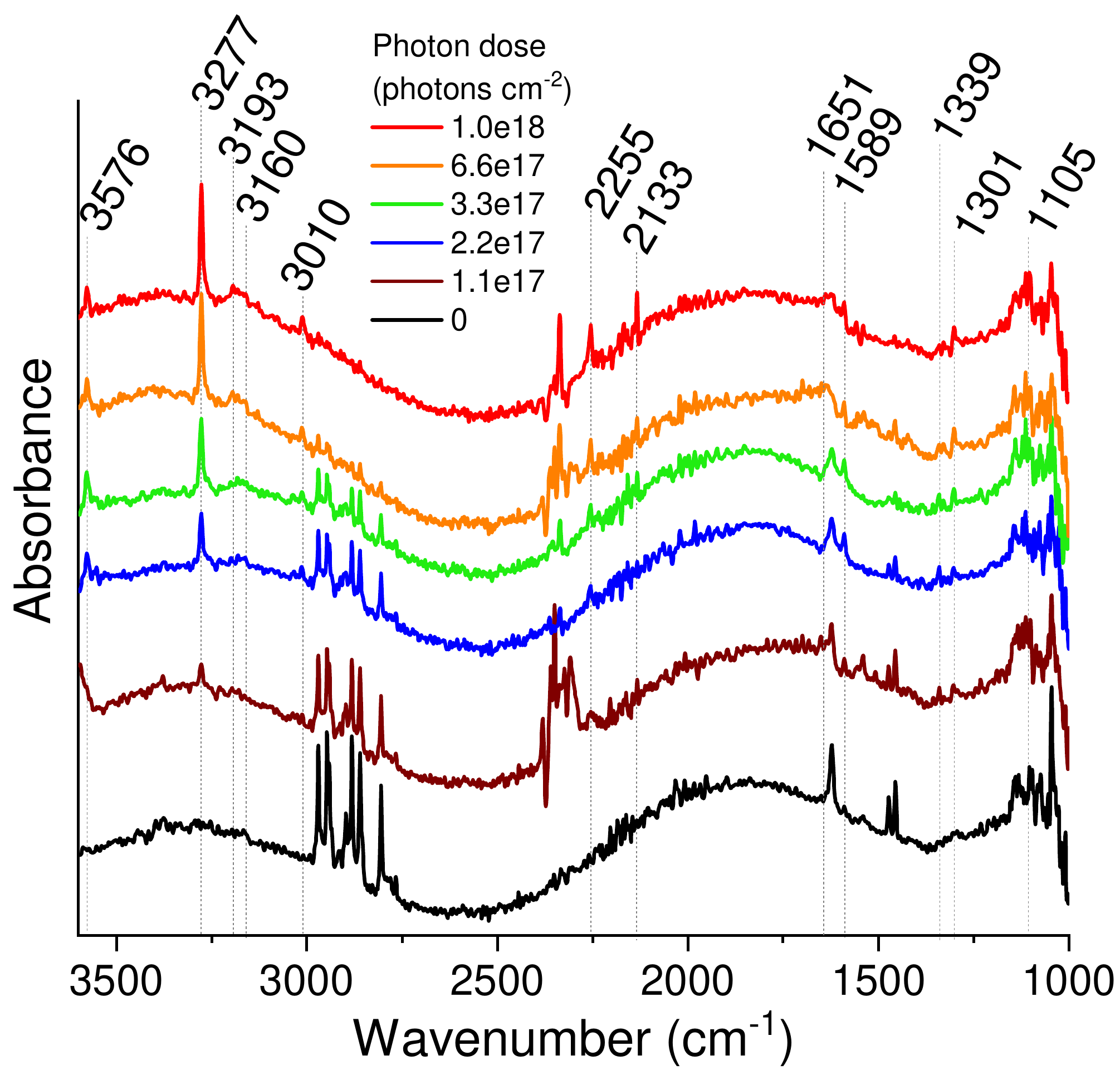}
  \caption{CH$_3$NH$_2$:Xe (Exp. \textbf{5}).}
  \label{Fig.CH3NH2-Xe-irradiacion}
\end{subfigure} 
\begin{subfigure}{.45\textwidth}
  \centering 
  \includegraphics[width=\linewidth]{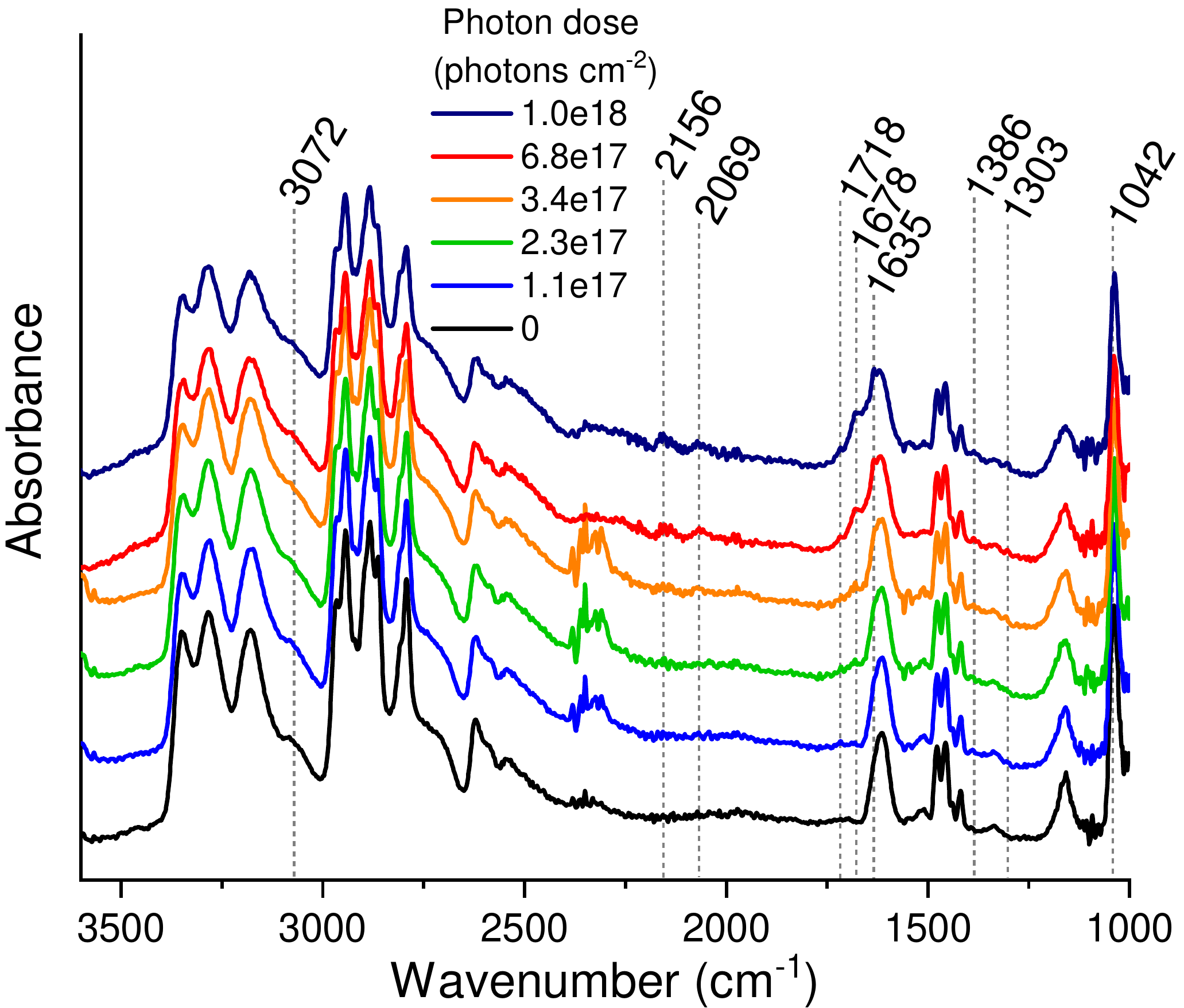}
  \caption{CH$_3$NH$_2$ (Exp. \textbf{2})}
  \label{Fig.CH3NH2_irradiacion}
\end{subfigure} 
\begin{subfigure}{.45\textwidth}
  \includegraphics[width=\linewidth]{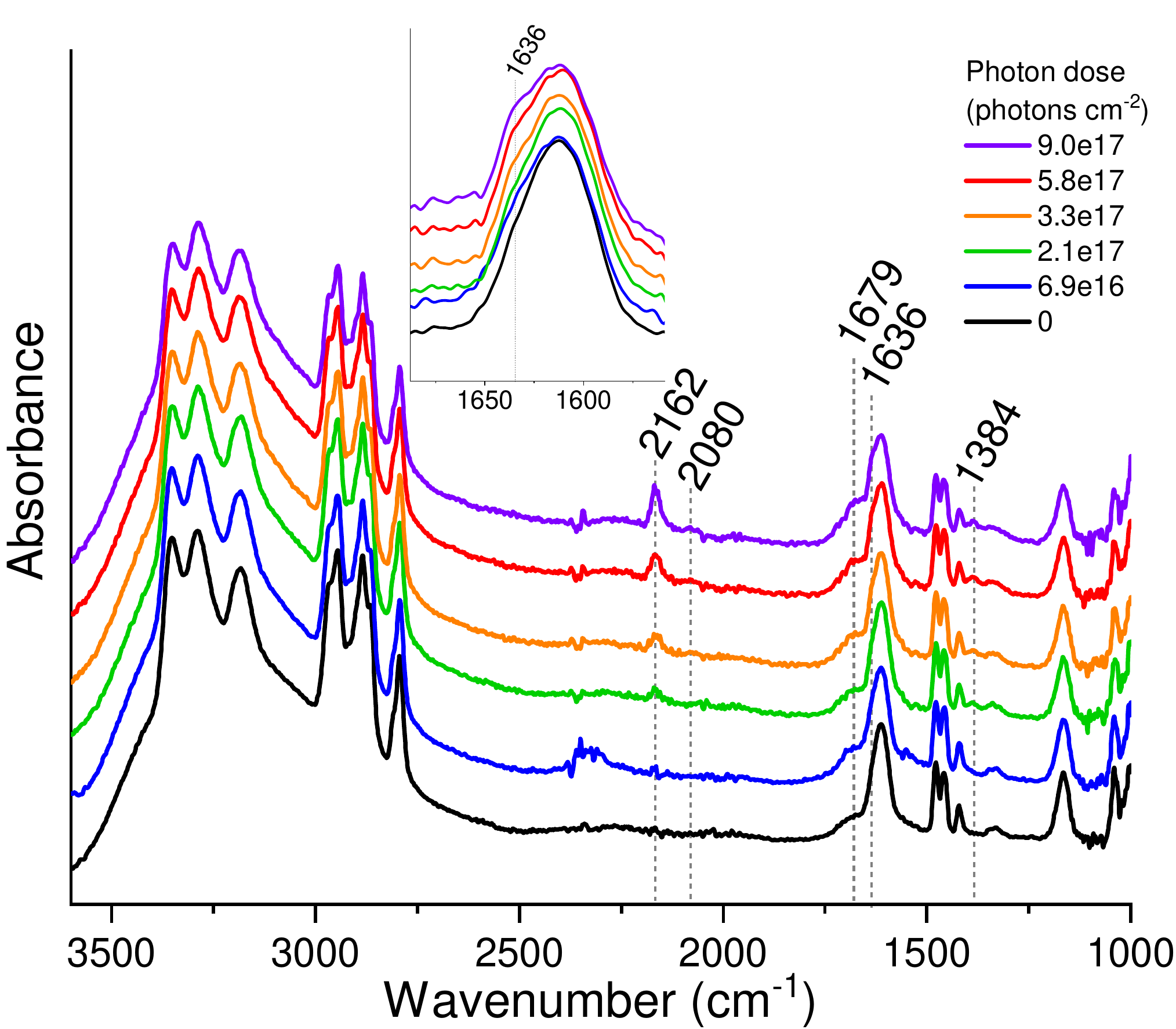}
  \caption{CH$_3$NH$_2$:H$_2$O (Exp. \textbf{6})}
  \label{Fig.CH3NH2_H2O_irradiacion}
\end{subfigure} 
\caption{Evolution of the IR spectra during the irradiation of three different ice samples. Vertical lines indicate the most relevant changes. The vibrational and molecular assignments are provided in Table \ref{Table.fotoproductos_irr}.}
\label{Fig.CH3NH2_UV_irradiation}
\end{figure} 

\subsection{UV irradiation of CH$_3$NH$_2$, CH$_3$NH$_2$:Xe and CH$_3$NH$_2$:H$_2$O ice samples. Photoproducts formation.}
\label{sect.fotoproductos}

\begin{table*} 
\Large
   \centering
    \caption[]{New IR features (in cm$^{-1}$) arising in the sample spectra during irradiation of the different ice mixtures at 8 K.}
    \label{Table.fotoproductos_irr}
    \resizebox{17cm}{!} {
    \begin{tabular}{lllllll}
Mass &Species & Vibration mode & CH$_3$NH$_2$:Xe & CH$_3$NH$_2$ (5\% H$_2$O) & CH$_3$NH$_2$:H$_2$O & Reference \\
\noalign{\smallskip}
\noalign{\smallskip}
\hline
\hline
\noalign{\smallskip}
16& CH$_4$               & C-H str.                    &3010             &3013             &3005 &\cite{d'Hendecourt1986AA...158..119D}\\
\noalign{\smallskip}
16& CH$_4$               & C-H bend.                   &1301             &1303             &1306 &\cite{d'Hendecourt1986AA...158..119D}\\
\noalign{\smallskip}
16& NH$_2\cdot$          & N-H bend.                   &1500             & -               &- &\cite{Bossa2012_ch3nh2_VUV}\\
\noalign{\smallskip}
17& NH$_3$               & N-H str.                    & -               &3390             &3405  &\cite{Ferraro1980ApSpe..34..525F, Danger2011AA...525A..30D}\\
\noalign{\smallskip}
17& NH$_3$               & umbrella mode               & -               & 1070            &1076  &\cite{Ferraro1980ApSpe..34..525F, Danger2011AA...525A..30D}\\
\noalign{\smallskip}
26& CN$^{-}$             & C$\equiv$N str.             & 2084            & 2069            &2080  &\cite{Moore2003, Danger2011AA...535A..47D}\\
\noalign{\smallskip}
27& HNC                  & N-H str.                    &3576             & -               &-  &\cite{Jacox1975333, Moore2003, mencos2018MNRAS.476.5432M}\\
\noalign{\smallskip}
27& HCN                  & C-H str.                    &3277             & -               &-  &\cite{Jacox1975333, Moore2003, mencos2018MNRAS.476.5432M}\\
\noalign{\smallskip}
27& HCN                  & C$\equiv$N str.             &2133             & -               &-  &\cite{Moore2003, mencos2018MNRAS.476.5432M}\\
\noalign{\smallskip}
29& CH$_2$NH             & CH$_2$ degenerate str.      & 3193, 3160      & -               &-  &\cite{Zhu2019PCCP...21.1952Z}\\
\noalign{\smallskip}
29& CH$_2$NH             & comb.                       & 2219            & -               &-  &\cite{Danger2011AA...535A..47D}\\
\noalign{\smallskip}
29& *CH$_2$NH            & C=N str.                    & 1651            & -               &-  &\cite{Jacox1975333, Danger2011AA...535A..47D, Zhu2019PCCP...21.1952Z}\\
\noalign{\smallskip}
29& CH$_2$NH             & C-N torsion                 & 1105            & 1119            &-  &\cite{Jacox1975333}\\
\noalign{\smallskip}
29& CH$_2$NH             &HCNH def.                    & 1059            & 1068            &-  &\cite{Jacox1975333}\\
\noalign{\smallskip}
30& H$_2$CO              & C-H str.                    & -               & 2828            & 2830 &\cite{Schutte1993Icar..104..118S, Vinogradoff2012}\\
\noalign{\smallskip}
30& H$_2$CO              & C=O str.                    & -               & 1718            & 1729 &\cite{Schutte1993Icar..104..118S, Vinogradoff2012}\\
\noalign{\smallskip}
30& H$_2$CO              & CH$_2$ bend.                & -               & 1498            &1493  &\cite{Schutte1993Icar..104..118S, Vinogradoff2012}\\
\noalign{\smallskip}
30& *$\cdot$CH$_2$NH$_2$ & N-H str.                    & -               & 3450-3380       &-  &\cite{Dyke1989IJMSI..94..221D}\\
\noalign{\smallskip}
30& CH$_3$NH$\cdot$       & comb.                      & 1025           & -               &- &\cite{Ruzi2012_radicals_IR}\\
\noalign{\smallskip}
32& CH$_3$NH$_3^+$       & comb.                       & -               & 2654            &2645  &\cite{Bossa2009AA...506..601B, Bossa2012_ch3nh2_VUV}\\
\noalign{\smallskip}
32& CH$_3$NH$_3^+$       & comb.                       & -               & 2553            &2552  &\cite{Bossa2009AA...506..601B, Bossa2012_ch3nh2_VUV}\\ 
\noalign{\smallskip}
41& CH$_3$CN             & C$\equiv$N str.             & 2255            & 2254            &-  &\cite{wexler1967ApSRv...1...29W, d'Hendecourt1986AA...158..119D, Danger2011AA...525A..30D}\\
\noalign{\smallskip}
42& OCN$^{-}$            & C$\equiv$N str.             & -               & 2156            &2162  &\cite{Moore2003, Jimenez2014ApJ...788...19J}\\
\noalign{\smallskip}
43& CH$_3$CHNH           & C=N str.                    & -               & 1668            &1679  &\cite{Stolkin1977CP.....21..327S, Danger2011AA...525A..30D}\\
\noalign{\smallskip}
43& CH$_3$CHNH           & CH$_3$ bend.                & -               & 1438            &1442  &\cite{Stolkin1977CP.....21..327S, Danger2011AA...525A..30D}\\
\noalign{\smallskip}
45& HCONH$_2$            & NH$_2$ scissoring           & -               & 1635            &1636 &\cite{Sivaraman2013AcSpA.105..238S}\\
\noalign{\smallskip}
45& HCONH$_2$            & CH bend.                    & -               & 1386            &1384  &\cite{Sivaraman2013AcSpA.105..238S}\\
\noalign{\smallskip}
45& CH$_3$CH$_2$NH$_2$   & N-H str.                    & -               & 3308            &3323  &\cite{Danger2011AA...525A..30D}\\
\noalign{\smallskip}
45& CH$_3$CH$_2$NH$_2$   & N-H str.                    & -               & 3219            &3236  &\cite{Danger2011AA...525A..30D}\\
\noalign{\smallskip}
45& CH$_3$CH$_2$NH$_2$   & N-H str.                    & -               & 3104            &3121  &\cite{Danger2011AA...525A..30D}\\
\noalign{\smallskip}
45& CH$_3$CH$_2$NH$_2$   & C-C-N str.                  & -               & 1095            &-  &\cite{Durig2006JPCA..110.5674D, Danger2011AA...525A..30D}\\
\noalign{\smallskip}
  & ?                     &                            & 1339            & -               & -  &\cite{}\\
\noalign{\smallskip}
  & ?                     &                            & 1589            & -               &-  &\cite{}\\
\noalign{\smallskip}
\hline
\end{tabular}
}
\textit{\\
\small
str. = stretching; bend. = bending; def. = deformation\\
* tentatively assigned.\\
}
\end{table*}

\begin{figure*}
  \centering 
  \includegraphics[width=1\textwidth]{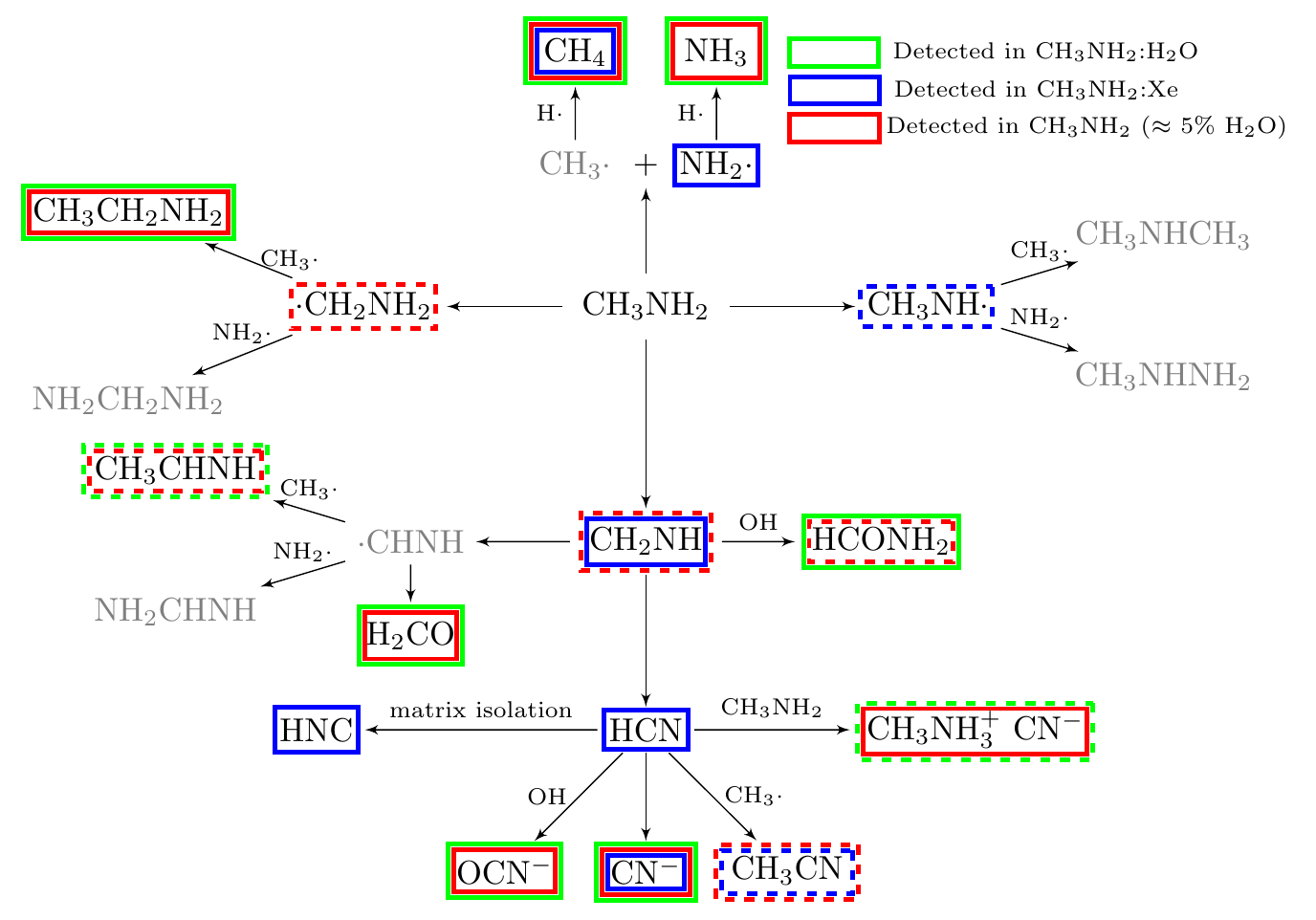}
  \caption{Species arising from UV irradiation at 8 K of different ice samples containing CH$_3$NH$_2$. Colours are indicative of species found in CH$_3$NH$_2$:H$_2$O = 100:5 (red), CH$_3$NH$_2$:Xe (blue), and CH$_3$NH$_2$:H$_2$O = 1:1 (green) ice samples. Light grey denote non-detected species, which require the reaction of, at least, one NH$_x$ group, less reactive than methyl groups, see Sect. \ref{Sect.H2_diffussion}. Dashed lines are indicative of a tentative detection, from reproducible bands which are close to the detection limit. Identification of the different species was made by the following IR features, some were also detected during thermal desorption by QMS: CH$_4$, bands at 1303 and 3013 cm$^{-1}$; NH$_3$, 3390 cm$^{-1}$; NH$_2\cdot$, decay of the N-H bending mode around 1500 cm$^{-1}$; $\cdot$CH$_2$NH$_2$, decay of N-H stretching around 3400 cm$^{-1}$; CH$_3$CH$_2$NH$_2$, bands near 3308, 3219, 3104, and 1095 cm$^{-1}$ after subtraction between irradiated and non-irradiated ice samples; CH$_3$NH$\cdot$, 1025 cm$^{-1}$ feature in Xe matrix; CH$_2$NH, bands at 3193, 3160, 2219, 1105, and 1059 cm$^{-1}$; HCONH$_2$, 1635 and 1386 cm$^{-1}$ and thermal desorption at 205 K ($\frac{m}{z} = 45$); H$_2$CO, bands at 2828 and 1718 cm$^{-1}$; CH$_3$CHNH, bands at 1668, 1438 cm$^{-1}$, and thermal desorption at 200 K ($\frac{m}{z} = 45$); HCN, bands at 3277, and 2133 cm$^{-1}$, thermal desorption at 114 K $\frac{m}{z} = 27$); HNC, 3576 cm$^{-1}$ band and thermal desorption at 105 K ($\frac{m}{z} = 27$); OCN$^{-}$, 2156 cm$^{-1}$; CN$^{-}$, 2069 cm$^{-1}$; CH$_3$CN, 2254 cm$^{-1}$; CH$_3$NH$_3^+$ CN$^{-}$, 2654, 2553, and 2069 cm$^{-1}$.}
  \label{Fig.esquema_reaccion_metilamina}
\end{figure*} 

\setcounter{equation}{0} 
\renewcommand{\theequation}{\Roman{equation}} 

IR spectra measured during the irradiation period of CH$_3$NH$_2$, CH$_3$NH$_2$:Xe and CH$_3$NH$_2$:H$_2$O ice samples (Exps. \textbf{2}, \textbf{6} and \textbf{7}) are shown in Fig. \ref{Fig.CH3NH2_UV_irradiation}, and Table \ref{Table.fotoproductos_irr} exhibits the assignment of the IR bands. Fig. \ref{Fig.esquema_reaccion_metilamina} summarizes the different species arising from UV irradiation at 8 K. The formation of different species at low temperature is presented starting from the simplest ice sample, CH$_3$NH$_2$:Xe to the most astrophysically relevant one, CH$_3$NH$_2$:H$_2$O. Pure CH$_3$NH$_2$ ice sample served as an intermediate step.\\

\subsubsection{UV irradiation of CH$_3$NH$_2$:Xe ice samples}

IR spectra during the irradiation of CH$_3$NH$_2$:Xe ice samples is shown in Fig. \ref{Fig.CH3NH2-Xe-irradiacion}. Matrix isolation of methylamine enhances the detection of radical species. $\cdot$NH$_2$ was detected by its IR feature at 1500 cm$^{-1}$, in line with the position of this radical formed from UV irradiation of NH$_3$ \citep{MD2018}. The large hydrogen subtraction hampers the formation of hydrogenated species. NH$_3$ was, in fact, not detected at 8 K, although the low N-H stretching band strength at low temperature may hinder its detection. CH$_4$, which can be formed by direct decomposition of methylamine molecules at wavelengths shorter than 388 nm \citep{Gardner1982}, was detected as a minor photoproduct. Furthermore, CH$_3\cdot$ radicals present in the CH$_3$NH$_2$:Xe ice mixture can only react with H$\cdot$ radicals of H$_2$ molecules, as they are the only species able to move fast enough at 8 K, thus enhancing CH$_4$ formation.\\

According to \cite{Bossa2012_ch3nh2_VUV}, CH$_3$NH$\cdot$ and $\cdot$CH$_2$NH$_2$ radicals are also expected to form directly from methylamine upon UV irradiation. The former was detected by its 1025 cm$^{-1}$ IR feature, while the latter was not identified, probably as a consequence of its lower formation rate \citep{Bossa2012_ch3nh2_VUV}. Additionally, even if $\cdot$CH$_2$NH$_2$ is formed, it tends to react, either with $\cdot$CH$_3$ radicals to form ethylamine, CH$_3$CH$_2$NH$_2$, or loosing a hydrogen atom to form CH$_2$NH. Nitrogen atoms, however, can stabilize the electron, thus reducing the reactivity of R-NH$_x$ radicals.\\

UV irradiation of matrix isolated methylamine led, preferentially, to the formation of CH$_2$NH molecules by hydrogen subtraction from methylamine. As reported by \cite{Gardner1982}, the dissociation of CH$_3$NH$_2$ molecules producing CH$_2$NH and H$_2$ is the most energetically favoured process concerning UV irradiation of methylamine in the gas phase. Alternatively, CH$_2$NH can be obtained from hydrogen elimination from CH$_3$NH$\cdot$ and $\cdot$CH$_2$NH$_2$ radicals. Methylenimine was detected through its IR features at 3193, 3160, 2219, 1651, 1113, and 1059 cm$^{-1}$. CH$_2$NH molecules can also undergo hydrogen subtraction. Thus, HCN molecules are obtained. HCN was detected by its IR features at 3277 cm$^{-1}$ and 2133 cm$^{-1}$ (see Table \ref{Table.fotoproductos_irr}). HNC molecules, which are far less stable than its HCN isomer, were also detected by its IR feature at 3576 cm$^{-1}$ (Fig. \ref{Fig.CH3NH2-Xe-irradiacion}). As reported by \cite{Milligan1967JChPh..47..278M}, HNC formation is enhanced in a matrix environment. The absorption at 2084 cm$^{-1}$ is indicative of the presence of cyanide, CN$^{-}$, species, although it could not be attributed to any specific salt.\\

As depicted in Table \ref{Table.fotoproductos_irr}, UV irradiation of CH$_3$NH$_2$:Xe ice samples promotes the formation of small species, which can be produced from just one parent CH$_3$NH$_2$ molecule. CH$_3$CN, tentatively detected in isolated methylamine UV irradiation, was the only species that involves reaction of two CH$_3$NH$_2$ molecules. Indeed, CH$_3$CN was only detected in an advanced stage of the irradiation period, suggesting the reduced mobility of radicals at 8 K.\\ 

\subsubsection{UV irradiation of pure CH$_3$NH$_2$ ice samples} 

IR spectra of a pure CH$_3$NH$_2$ ice sample is represented in Fig. \ref{Fig.CH3NH2_irradiacion}. CH$_3$NH$\cdot$, $\cdot$CH$_2$NH$_2$, $\cdot$CH$_3$ and $\cdot$NH$_2$ radicals can be directly obtained from UV irradiation of methylamine \citep{Bossa2012_ch3nh2_VUV}. In the pure CH$_3$NH$_2$ ice sample, radicals bearing NH groups appear as broad bands at both sides of the N-H stretching region of the IR spectrum. In line with \cite{MD2018}, radicals appear readily within the irradiation period, but they are readily consumed, keeping a constant formation and destruction rate for longer irradiation times.\\

The lower hydrogen diffusion in the pure CH$_3$NH$_2$ ice, compared to matrix isolated methylamine, is expected to produce a larger fraction of hydrogenated species, such as NH$_3$ and CH$_4$. NH$_3$ and CH$_4$ were, in fact, detected in CH$_3$NH$_2$ ice samples (see Table \ref{Table.fotoproductos_irr} and Fig. \ref{Fig.CH3NH2_irradiacion}). \cite{MD2018} demonstrated that NH$_2\cdot$ radicals can remain in a pure NH$_3$ ice until thermal energy enables chemical reactions. Indeed, NH$_2\cdot$ radicals can only react with hydrogen atoms at 8 K to form NH$_3$ molecules in methylamine ice, explaining the plateau reached in NH$_2\cdot$ formation in \cite{MD2018}. The low absorbance at 1303 and 3013 cm$^{-1}$ is indicative of a minor formation rate of CH$_4$ in pure CH$_3$NH$_2$ ice samples. Irradiation of pure methane ice also produces CH$_3\cdot$ radicals, which react very readily to form small hydrocarbons and hydrogenated amorphous carbon, even at 8 K \citep[][and references therein]{Carrascosa2020MNRAS.493..821C}. As it will be explained later on, IR spectra from pure CH$_3$NH$_2$ ice displays ethylamine (CH$_3$CH$_2$NH$_2$), ethylenimine (CH$_3$CHNH) and acetonitrile (CH$_3$CN) IR features. The formation of many CH$_3$- bearing species suggests that CH$_3\cdot$ radicals are more prone to react with other species rather than H$\cdot$ radicals, as it occurs in the CH$_4$ ice UV irradiation experiments \citep{Carrascosa2020MNRAS.493..821C}. Despite the fast formation of ethane and propane, these authors showed that the reaction CH$_3\cdot$ + H$\cdot$ $\rightarrow \; $CH$_4$ is not efficient, at least in the ice surface. As a consequence, the interaction of H$\cdot$ radicals in CH$_4$ ice is small and H$_2$ desorption is 100 times larger in a pure CH$_4$ ice than the one found in CH$_3$NH$_2$ ice in our experiments for similar ice thicknesses (Fig. \ref{Fig.comparison_H2_production}).\\

\begin{figure}
  \centering 
  \includegraphics[width=0.5\textwidth]{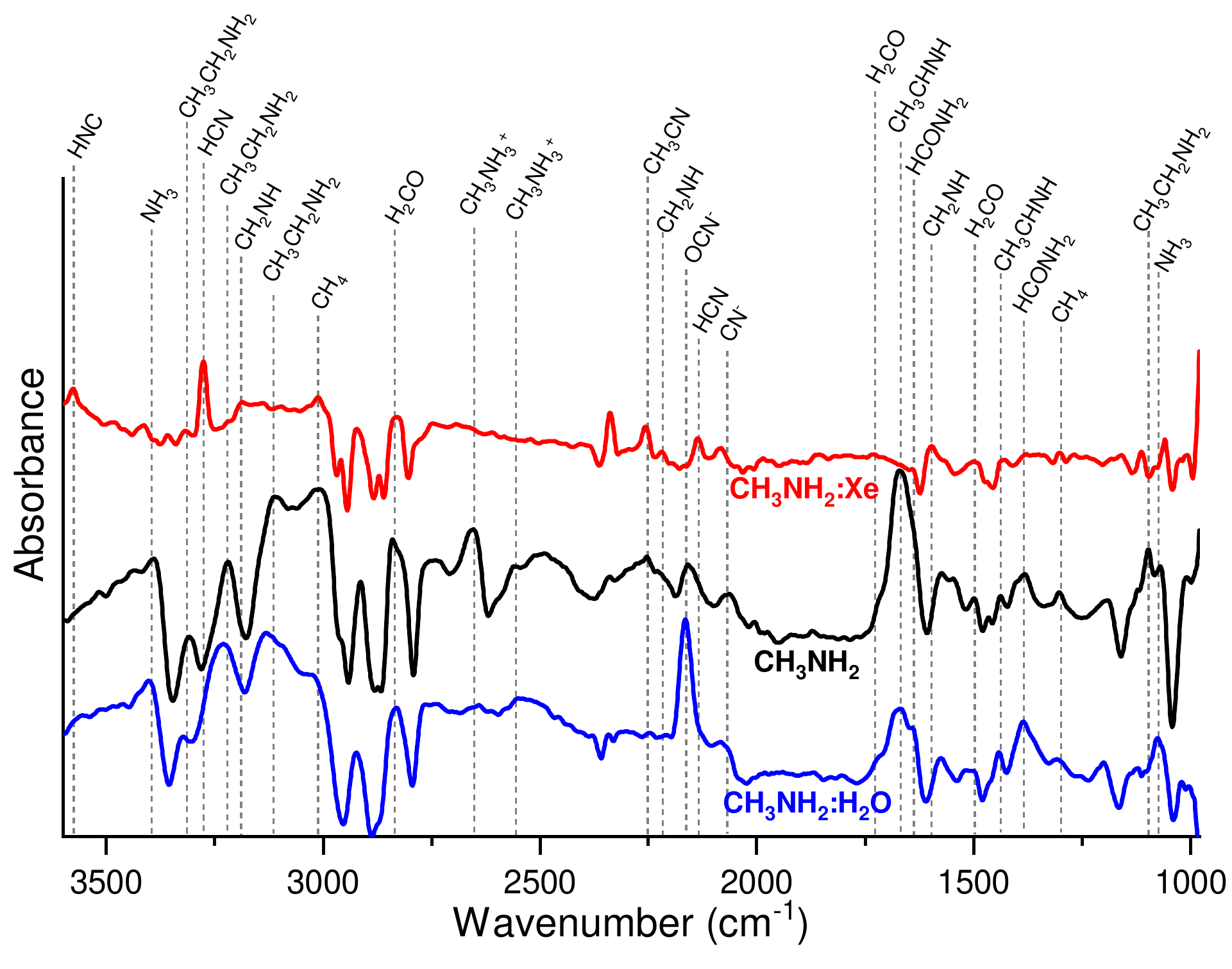}
  \caption{IR spectra of irradiated CH$_3$NH$_2$:Xe (red), CH$_3$NH$_2$ (black) and CH$_3$NH$_2$:H$_2$O (blue) ice mixtures after subtraction of the corresponding non-irradiated IR spectra.}
  \label{Fig.IR_difference}
\end{figure}

CH$_3\cdot$ and $\cdot$CH$_2$NH$_2$ radicals will react leading to an efficient formation of ethylamine. Ethylamine IR features overlap with those of methylamine. Yet, the relative intensity between the symmetric and the antisymmetric N-H stretching modes \citep{Hashiguchi1984JMoSp.105...81H, Danger2011AA...525A..30D} allowed us to infer its presence. The difference spectra between the irradiated and the non-irradiated methylamine ice (shown in Fig. \ref{Fig.IR_difference}) confirmed the presence of ethylamine by its IR features at 3308, 3219, 3104 and 1095 cm$^{-1}$ (see Table \ref{Table.fotoproductos_irr}).\\

CH$_2$NH was tentatively detected in pure CH$_3$NH$_2$ ice by its IR bands at 1119 and 1068 cm$^{-1}$. The presence of other molecules and radicals around CH$_2$NH molecules enhance its reactivity, lowering its abundance with respect to matrix isolated methylamine. Methylenimine reaction with CH$_3\cdot$ radicals produces ethylenimine (CH$_3$CHNH), which was detected by its IR features at 1438 cm$^{-1}$ and 1668 cm$^{-1}$. Dehydrogenation of CH$_3$CHNH molecules produces CH$_3$CN, responsible of the IR band at 2254 cm$^{-1}$. Dehydrogenation of CH$_2$NH molecules produces hydrogen cyanide, HCN. HCN molecules were not detected in pure CH$_3$NH$_2$ ice samples, but CN$^{-}$ anions were. As reported by \cite{Bossa2012_ch3nh2_VUV}, HCN and surrounding CH$_3$NH$_2$ molecules give rise to methylammonium cyanide [CH$_3$NH$_3^+$ CN$^{-}$] salt, detected in the IR spectra by the IR features at 2654 and 2553 cm$^{-1}$ from the cation and 2069 cm$^{-1}$ from the anion (see Table \ref{Table.fotoproductos_irr} and Fig. \ref{Fig.CH3NH2_irradiacion}). This reaction is favoured by the formation of CH$_3$NH$_3^+$ cation \citep{Zhang2017JPCA..121.7176Z}, which explains the lower formation rate of CN$^{-}$ in the CH$_3$NH$_2$:Xe experiment. The formation of other cyanides, however, cannot be excluded, with IR bands overlapping in the 2069 cm$^{-1}$ feature.\\

\begin{figure*}
  \centering 
  \includegraphics[width=\textwidth]{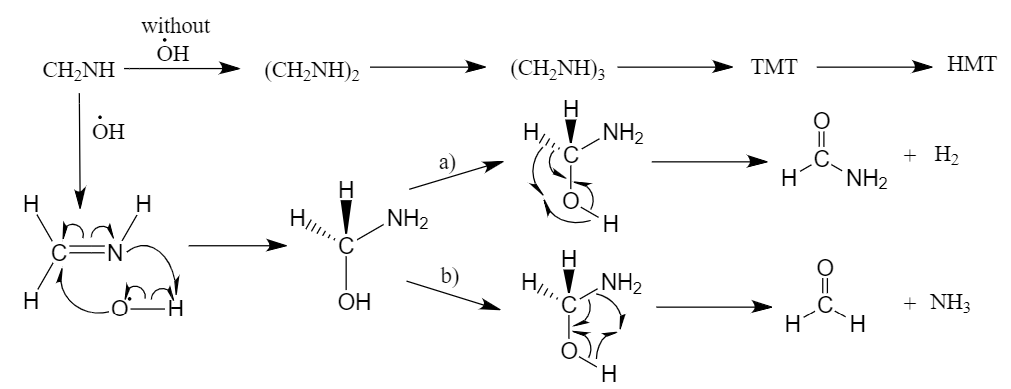}
  \caption{Chemical reactions of methylenimine molecules with and without surrounding OH radicals, formed from H$_2$O dissociation under UV photons. The presence/absence of water determines the preferential formation of formamide and formaldehyde, or HMT, respectively. Formamide, HCONH$_2$, and formaldehyde, H$_2$CO, can be formed at low temperature from radicalary reactions induced by UV photons \citep{bossa2009ApJ...707.1524B}.}
  \label{Fig.hidrolisis_imina}
\end{figure*}

When referring to pure methylamine ice, we assume that some H$_2$O molecules ($\approx$ 5\%) are present, due to the deposition process, as explained in Sect. \ref{sect.experimental_setup}. Although CH$_3$NH$_2$ vapour pressure is higher than that of H$_2$O, it is not possible to get rid of all the H$_2$O. Therefore, some oxygenated compounds were detected from UV irradiation of CH$_3$NH$_2$ ice. Chemical reactions between CH$_2$NH and OH radicals (formed from UV irradiation of H$_2$O molecules) are shown in Fig. \ref{Fig.hidrolisis_imina}. The nucleophylic attack of an OH radical over a carbon atom in a methylenimine molecule leads to the formation of aminomethanol as an intermediate. Formamide, HCONH$_2$, formation has been reported at low temperature during UV irradiation of aminomethanol ice through radical reactions \citep{bossa2009ApJ...707.1524B}. Additionally, the nucleophylic attack by the alcohol group over the aminomethanol intermediate results in formamide formation in a barrier-less reaction, as reported by \cite{Vazart2016_formamide_from_CH2NH_OH}. \cite{MD2020ApJ...894...98M} reported the formation of formamide from radical reaction between NH$_2$ and HCO radicals, although the relatively low formation rate of these radicals in CH$_3$NH$_2$:H$_2$O ice samples suggests a negligible formation through this mechanism in our experiments. Formamide was identified by its 1386 cm$^{-1}$ and 1635 cm$^{-1}$ IR features in Fig. \ref{Fig.CH3NH2_irradiacion} and Table \ref{Table.fotoproductos_irr} (note that the 1635 cm$^{-1}$ band is observed as a contribution to the left shoulder of the prominent band around 1610 cm$^{-1}$). These features display a weak intensity but they were highly reproducible in our experiments. As shown in Fig. \ref{Fig.hidrolisis_imina}, elimination of the amino group during the second nucleophylic attack results in formaldehyde, H$_2$CO, detected in the IR spectrum by its C=O stretching absorption at 1718 cm$^{-1}$, and ammonia formation \citep{Layer1963_chemistry_imines}. Furthermore, dehydration of formamide produces HNCO, which readily undergoes hydrogen subtraction in presence of CH$_3$NH$_2$ to produce OCN$^{-}$ anions, detected by its main IR feature at 2156 cm$^{-1}$, and CH$_3$NH$_3^+$ cations. The relatively large band strengths of oxygenated compounds facilitate their detection, even when H$_2$O is a minor compound in the ice sample.\\

\subsubsection{UV irradiation of pure CH$_3$NH$_2$:H$_2$O ice samples} 

The presence of methylamine was inferred in comet 67P/Churyumov-Gerasimenko and this species was detected in the gas phase toward molecular clouds and hot cores, see Sect. \ref{sect.introduction} \citep{Goesmann2015Sci...349b0689G}. Pure methylamine ice, however, is not a good analog of astrophysical ices. Thus, more astrophysically realistic H$_2$O:CH$_3$NH$_2$ ice experiments were performed. The understanding of the processes in CH$_3$NH$_2$:Xe and pure CH$_3$NH$_2$ ice samples was, however, found to be crucial to understand the processes going on in this binary ice mixture.\\

CH$_3$NH$_2$:H$_2$O ice mixtures were prepared at different ratios between the components (see Table \ref{Table.experiments}). Fig. \ref{Fig.CH3NH2_H2O_irradiacion} shows the evolution of the IR spectra of Exp. \textbf{6}, where a column density of 3.9$\times$10$^{17}$ cm$^{-2}$ and 3.5$\times$10$^{17}$ cm$^{-2}$ for CH$_3$NH$_2$ and H$_2$O, respectively, was measured.\\

The presence of OH groups raises the reactivity of radicals within the CH$_3$NH$_2$:H$_2$O ice sample, in comparison to pure CH$_3$NH$_2$ and CH$_3$NH$_2$:Xe, preventing from the detection of radical species. The formation of ethylamine, CH$_3$CH$_2$NH$_2$ (IR bands at 3323, 3236 and 3121 cm$^{-1}$) is indicative of the presence of $\cdot$CH$_2$NH$_2$ and $\cdot$CH$_3$ radicals, as it occurs in the pure CH$_3$NH$_2$ ice sample. IR bands at 3005 and 1306 cm$^{-1}$ (CH$_4$) and 3405 and 1076 cm$^{-1}$ (NH$_3$), confirm the enhanced formation of these species when hydrogen bonds are established.\\

Although the highly reactive CH$_2$NH was not identified, its presence as an intermediate species can be inferred by the presence of IR bands at 1636 and 1384 cm$^{-1}$ (related to HCONH$_2$, which are enhanced when compared to pure CH$_3$NH$_2$ ice samples), 1679 and 1142 cm$^{-1}$ (CH$_3$CHNH), and 2830, 1729 and 1493 cm$^{-1}$ (H$_2$CO). Dehydrogenation of CH$_2$NH produces HCN. HCN was not identified in CH$_3$NH$_2$:H$_2$O ice mixture. As it was observed in pure CH$_3$NH$_2$ ice samples, HCN tends to react, producing CN$^{-}$, and OCN$^{-}$ species, which were identified by their absorptions at 2080 and 2162 cm$^{-1}$, respectively. CH$_3$NH$_2$ molecules can accept hydrogen atoms from HCNO and HCN, to produce the corresponding methylammonium salts ([CH$_3$NH$_3^+$ CN$^{-}$] and [CH$_3$NH$_3^+$ OCN$^{-}$]). IR bands at 2645 and 2552 cm$^{-1}$ are indicative of the presence of CH$_3$NH$_3^+$.\\

As it could be expected, IR spectra shown in Fig. \ref{Fig.CH3NH2_irradiacion} and Fig. \ref{Fig.CH3NH2_H2O_irradiacion} show that the presence of H$_2$O diminishes CH$_3$CN and CH$_3$CH$_2$NH$_2$ formation, while the formation of oxygenated compounds, such as OCN$^{-}$ and HCONH$_2$ is enhanced.\\

\subsection{Irradiation of H$_2$O:CH$_3$OH:NH$_3$ ice analogs}
We performed another experiment for comparison of the residue spectra obtained from methylamine-bearing ices with the better known residue made by irradiation and warm-up of the H$_2$O:CH$_3$OH:NH$_3$ (20:1:1) ice mixture. This experiment was discussed in previous works \citep[i. e.][]{Bernstein1995, Guille2003AA...412..121M}, and in this work, we only focus on the residue. This experiment served to check the formation of a "classical" residue under the UHV conditions of ISAC. The IR spectrum of this residue, presented in Fig. \ref{Fig.HMT_formation_comparison}, is similar to the one reported in previous works.\\

\subsection{Synthesis of large organics during warming up of UV irradiated CH$_3$NH$_2$ ice samples.}
\label{sect.TMT}

\begin{figure}
  \centering 
  \includegraphics[width=0.5\textwidth]{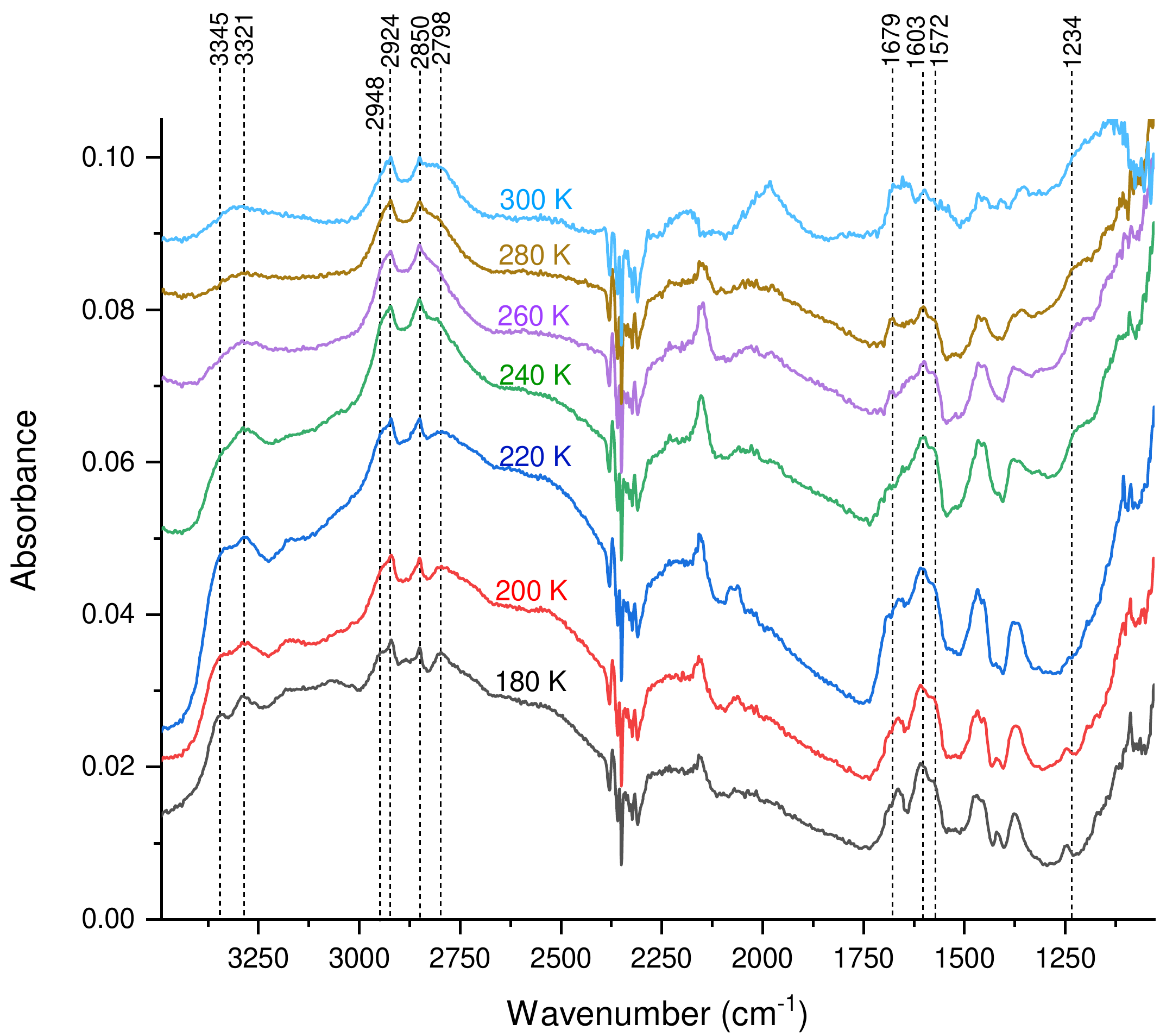}
  \caption{Evolution of the IR spectrum during warming up of a pure CH$_3$NH$_2$ ice sample (Exp. \textbf{4}).}
  \label{Fig.IR_TPD_residuo}
\end{figure} 

The reduced mobility of most species at the low temperatures present in interstellar ice mantles (10-20 K), and reproduced in our experiments, inhibits the formation of large species prior to the warm-up phase. Furthermore, the relatively low column density values of methylamine ices from Exps. \textbf{2}, \textbf{6} and \textbf{7} (see Table \ref{Table.experiments}) prevented us from the detection of larger organic molecules. The irradiation of thicker methylamine ices gives us the possibility to detect larger species with our IR spectrometer. Fig. \ref{Fig.IR_TPD_residuo} shows IR spectra recorded after thermal desorption of methylamine molecules, from 180 K to 300 K. Bands located at 3345 and 3321 cm$^{-1}$ are related to the symmetric and antisymmetric modes of NH$_2$ groups, that is, primary amines, at 180 K. As the temperature increases, both vibration modes are merged, becoming indistinguishable above 240 K. The same trend is observed for CH$_3$ groups, as the band at 2948 cm$^{-1}$ dissapears at the same temperature. NH$_2$ and CH$_3$ groups can be only bonded to one group, in other words, they are located at the end of the molecules or in branched species. The absence of these groups at larger temperatures is originated by two processes: thermal desorption of species to the gas phase, and the formation of larger chains and cyclic compounds that are refractory at those temperatures. Bands located in the 1575-1625 cm$^{-1}$ range, which remain at 300 K, may be indicative of the presence of cyclic aromatic compounds.\\

\begin{figure*}
\begin{subfigure}{.46\textwidth}
  \centering 
  \includegraphics[width=\linewidth]{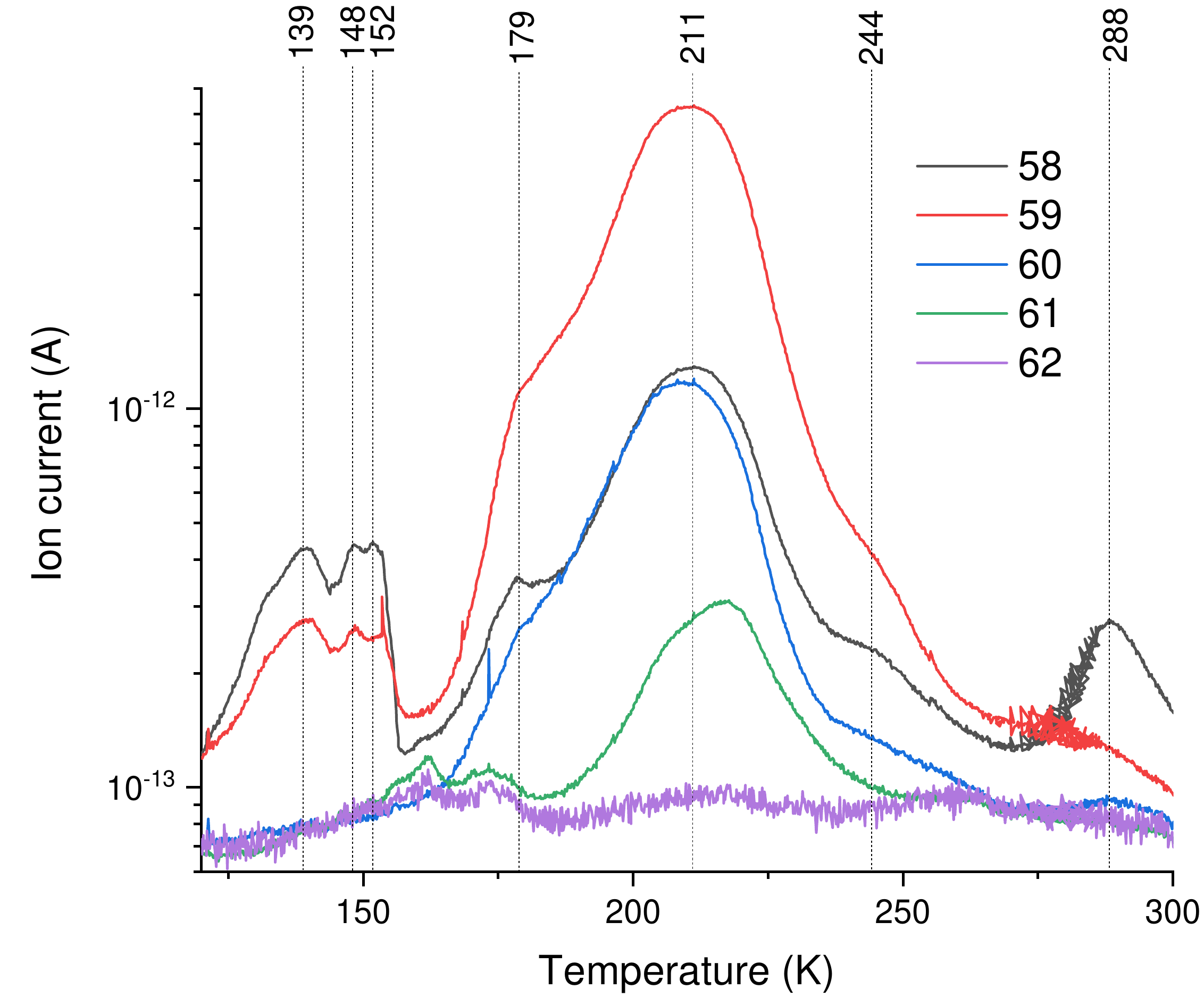}
  \label{Fig.QMS_TMT_fragments}
\end{subfigure} 
\begin{subfigure}{.53\textwidth}
  \includegraphics[width=\linewidth]{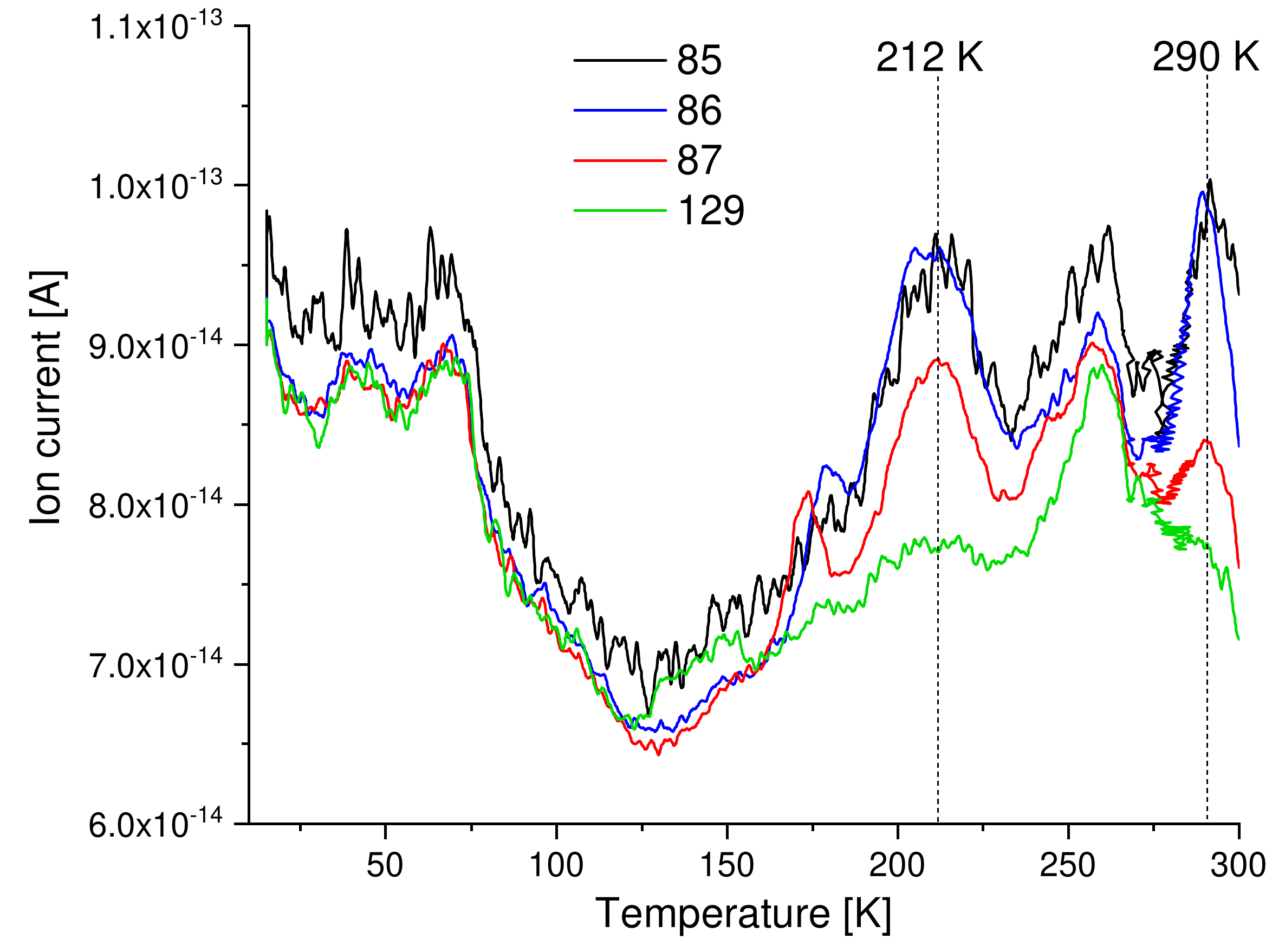}
  \label{Fig.QMS_TMT}
\end{subfigure} 
\caption{Warming up of an irradiated CH$_3$NH$_2$ ice sample. Left: recorded QMS signal for $\frac{m}{z}= 58, 59, 60, 61$ and 62. These $\frac{m}{z}$ fragments are representative of methylamine dimers. The molecular ion of non-covalent and covalent methylamine dimers has $\frac{m}{z} = 62$ and 60, respectively. Right: recorded QMS signal for molecular ion of TMT ($\frac{m}{z} = 87$) as well as its corresponding M-1 and M-2 fragments (86 and 85, see text). No molecule is expected to desorb with a $\frac{m}{z}$ ratio of 129, it was used for reference.}
  \label{Fig.QMS_TMT_dimers}
\end{figure*}

Three different methylamine dimers can be formed from UV irradiation of methylamine, CH$_3$-NH-CH$_2$-NH$_2$ (\textbf{Dimer A}), CH$_3$-NH-NH-CH$_3$ (\textbf{Dimer B}), and NH$_2$-CH$_2$-CH$_2$-NH$_2$ (\textbf{Dimer C}). The three dimers have the same molecular mass, hampering its unequivocal identification. Therefore, the most likely interpretation of the results is discussed in this work, although other species may contribute to the measured ion current of each of the $\frac{m}{z}$ values. Fig. \ref{Fig.QMS_TMT_dimers} suggests the presence of the three dimers in Exp. \textbf{8}. From 125 K, the ion current measured for $\frac{m}{z} = 58$ and 59 could be related to the desorption of \textbf{Dimer B}. The presence of methyl groups at both ends decreases the capability of the central NH groups to form hydrogen bonds due to steric interactions, thus reducing their thermal desorption temperature. In fact, although the molecular mass of \textbf{Dimer B} doubles that of methylamine, the absence of hydrogen bonds in \textbf{Dimer B} determines the codesorption of both species. It is not surprising that hydrogen bonds highly modify the intermolecular interactions. As an example the high desorption temperature of pure NH$_3$ ice, 95 K, compared to pure CH$_4$ ice, 40 K, is due to H bonds in NH$_3$ ice \citep[e.g.][]{Carrascosa2020MNRAS.493..821C, MD2018}.\\

\begin{figure*}
\begin{subfigure}{.46\textwidth}
  \centering 
  \includegraphics[width=\linewidth]{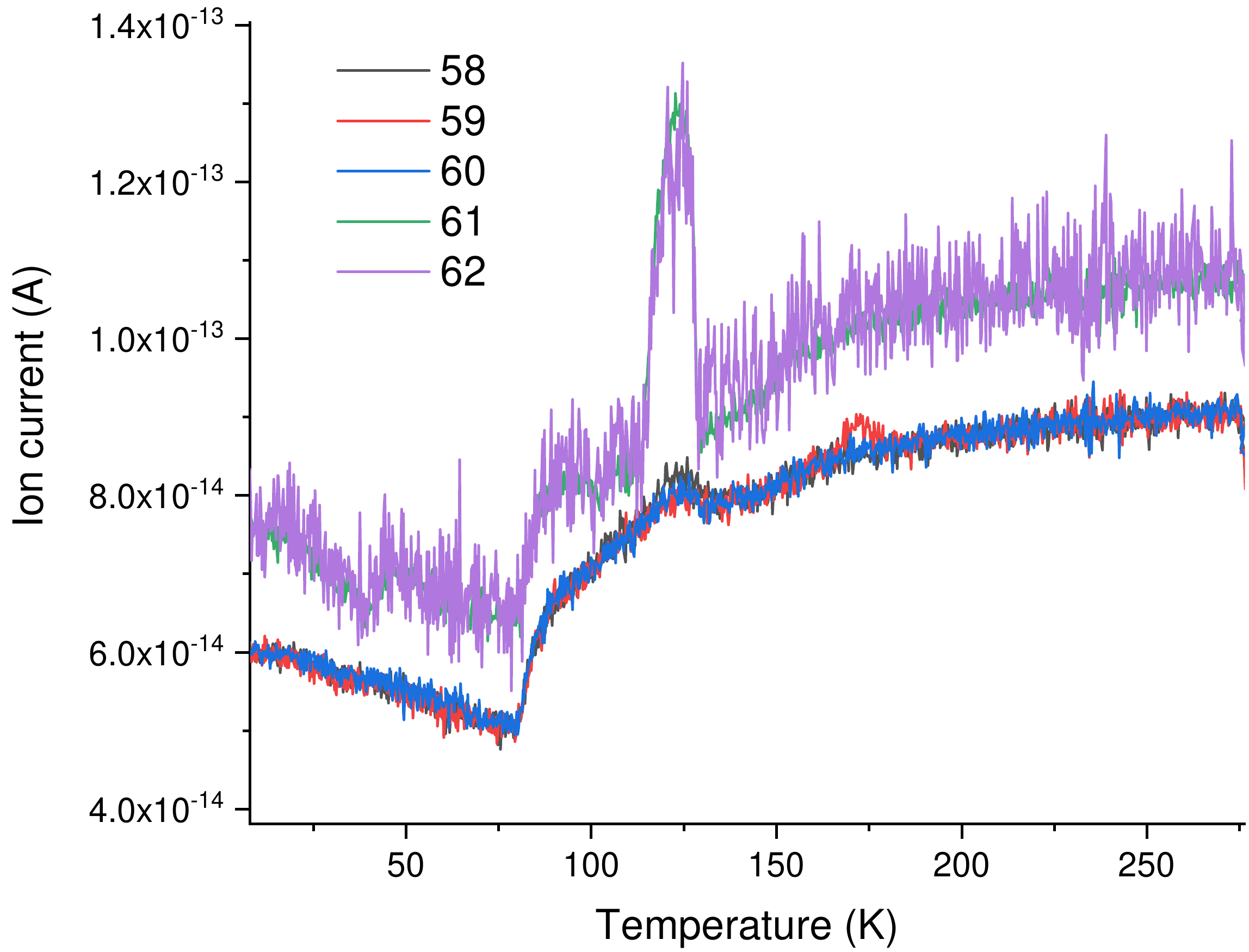}
  \label{Fig.QMS_TMT_fragments}
\end{subfigure} 
\begin{subfigure}{.53\textwidth}
  \includegraphics[width=\linewidth]{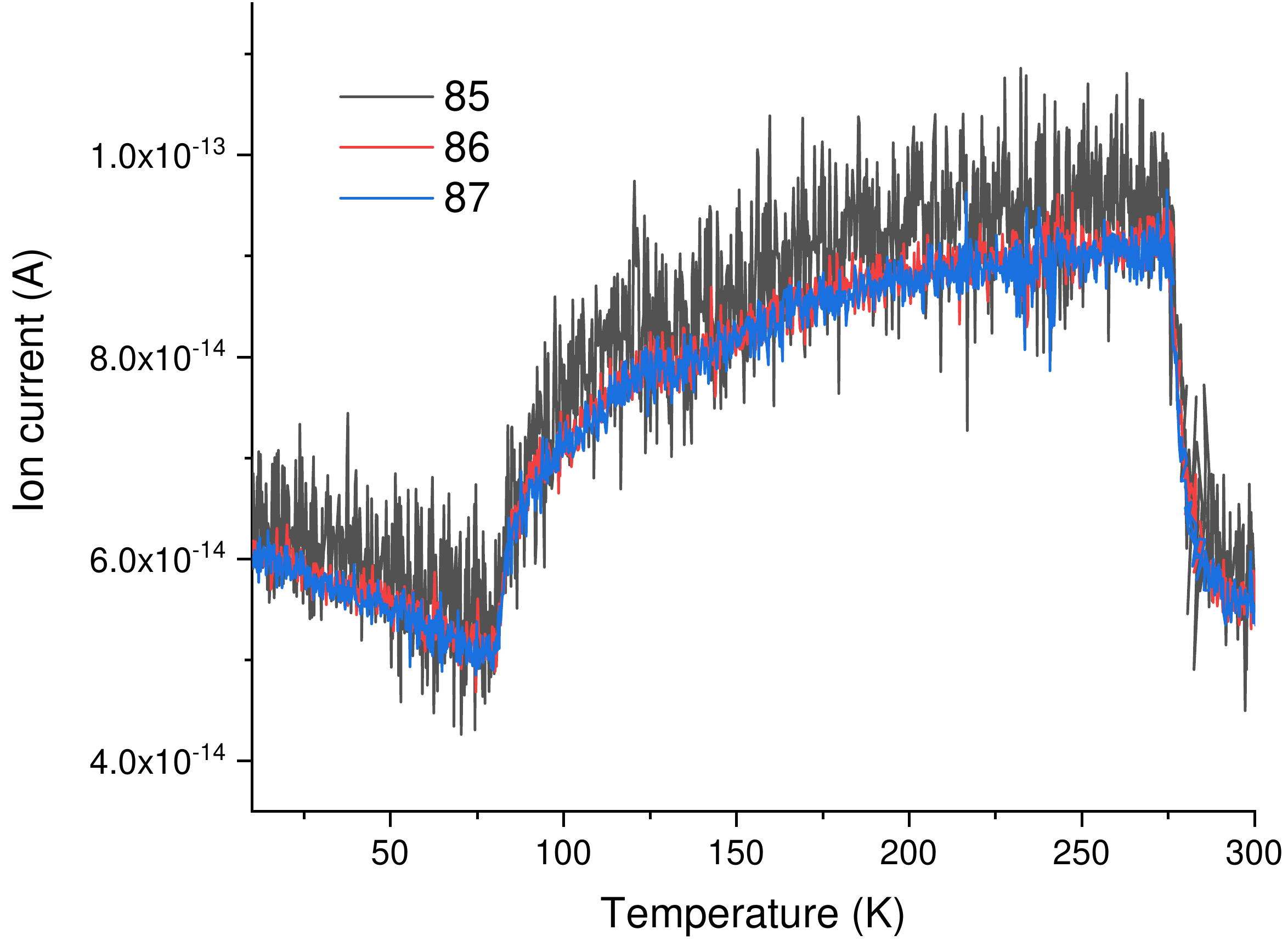}
  \label{Fig.QMS_TMT}
\end{subfigure} 
\caption{Warming up of a pure CH$_3$NH$_2$ ice without any irradiation period. Left: recorded QMS signal for $\frac{m}{z}= 58, 59, 60, 61$ and 62, representing $\frac{m}{z}$ ratios for methylamine dimers, being the non-covalent dimer ($\frac{m}{z} = 62$) the most abundant species. Right: recorded QMS signal for $\frac{m}{z}= 85, 86, 87$ in the same experiment, where no TMT formation is expected, in contrast to Fig. \ref{Fig.QMS_TMT_dimers}.}
  \label{Fig.QMS_TMT_blanco}
\end{figure*} 

\begin{figure}
  \centering 
  \includegraphics[width=0.5\textwidth]{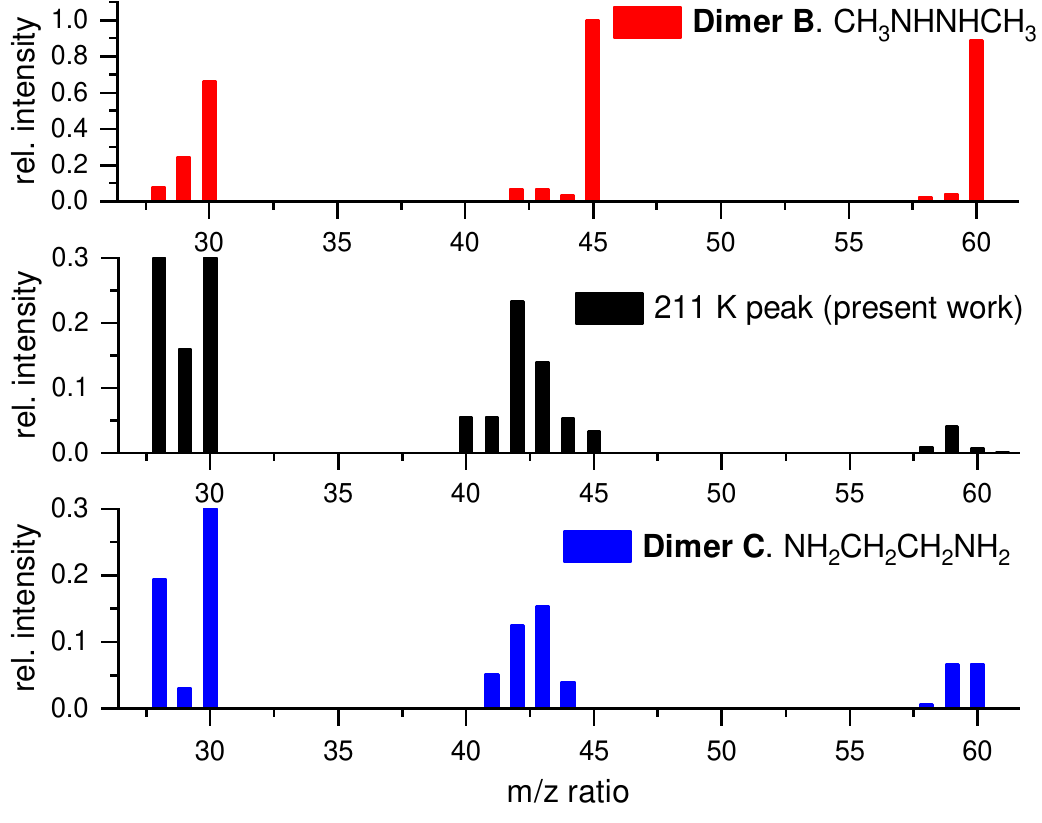}
  \caption{Mass spectra of \textbf{Dimer B} (top) and \textbf{Dimer C} (bottom) taken from NIST database. Relative intensities measured in our experiments during the desorption of the 211 K peak shown in Fig. \ref{Fig.QMS_TMT_dimers} are shown for comparison (middle). To our knowledge, the mass spectrum of \textbf{Dimer A} has not been published. \textbf{Dimer A} might contribute to $\frac{m}{z}$ = 28, 29, 40 and 42. $\frac{m}{z}$ = 28 and 30 have roughly the same intensity in the middle spectrum.}
  \label{Fig.mass_spectra}
\end{figure}

\textbf{Dimer C} is expected to be formed preferentially over the other dimers via reaction NH$_2$CH$_2\cdot$ + NH$_2$CH$_2\cdot$, as methyl groups are more reactive at low temperatures than NH groups (see Sect. \ref{sect.fotoproductos}). The two NH$_2$ groups at both ends enhances the formation of hydrogen bonds, thus retarding thermal desorption of \textbf{Dimer C} to 211 K. Furthermore, the mass spectra of \textbf{Dimer B} and \textbf{Dimer C} are compared in Fig. \ref{Fig.mass_spectra} to the one measured for the 211 K peak with our QMS. Its profile highly differs from \textbf{Dimer B}, in particular the low intensities of $\frac{m}{z} = 45$ and 60. On the contrary, the measured mass spectrum is more similar to \textbf{Dimer C}. It is not clear if the differences observed between the experimental data and the reported mass spectrum of \textbf{Dimer C} are due to variation in the fragmentation pattern due to the use of different instruments (i. e. NIST database versus our QMS), or whether there is contribution of a codesorbing species in our experiment, such as \textbf{Dimer A}. Nevertheless, mass spectrum of \textbf{Dimer B} shows a preferential rupture of methyl groups, providing an intense ion current for $\frac{m}{z} = 45$. $\frac{m}{z} = 45$ in the mass spectrum at 211 K in our experiments is marginal, suggesting that desorbing species do not contain methyl groups.\\

As it will be discussed later on, thermal desorption of TMT was observed at 290 K (see Fig. \ref{Fig.QMS_TMT_dimers}). TMT is likely formed from the reaction between three methylenimine units. Desorption of TMT molecules is therefore indicative of a previous formation of \textbf{Dimer A}, and its subsequent reaction to form the more stable 6-member cyclic TMT. Reactivity of the three dimers is then explained by the preferential reactivity of CH$_x$ radicals when compared to NH$_x$ radicals. Reactivity between two CH$_2$ groups of NH$_2$CH$_2\cdot$ radicals produces \textbf{Dimer C}. However, \textbf{Dimer C} cannot further react, as both ends contain NH$_2$ groups. \textbf{Dimer B} formation is hampered, as it requires two NH groups to react and they react at temperatures higher than CH$_x$ radicals. \textbf{Dimer A} is thus the only dimer able to form at relatively high ratios, and to further react to produce larger species.\\

TPD data in Fig. \ref{Fig.QMS_TMT_dimers} suggests the formation of TMT molecules and its thermal desorption at 290 K. Signals recorded for $\frac{m}{z}$ ratios larger than 87 did not show any increase at 290 K. Thus, $\frac{m}{z} = 87$, should be the molecular ion (M$^{+}$), which coincides with the molecular mass of TMT. Despite the low ion current measured for $\frac{m}{z}$ ratios in right panel of Fig. \ref{Fig.QMS_TMT_dimers}, the absence of any signal at 290 K for larger $\frac{m}{z}$ ratios (e. g. $\frac{m}{z} = 129$, shown in Fig. \ref{Fig.QMS_TMT_dimers}), together with the blank experiment (Fig. \ref{Fig.QMS_TMT_blanco}), ensure the reliability of the measured ion current. The most intense fragment is obtained for $\frac{m}{z}$ = 86 (M$^{+}$-1), followed by $\frac{m}{z}$ = 85 (M$^{+}$-2). \cite{Vinogradoff2012} and references therein reported polymerization of CH$_2$-NH molecules to produce polymethylenimine (PMI, (-CH$_2$NH-)$_n$). No IR bands related to PMI, however, were detected in our experiments. Instead, the molecular ion, $\frac{m}{z} = 87$, suggests a cyclic species formed by three CH$_2$-NH units. A linear molecule formed from CH$_2$NH polymerization will imply very reactive CH$_2$ and NH ends, and such molecules will further polymerize, or incorporate CH$_3$/NH$_2$ ends, thus changing the mass of the molecular ion. Finally, the presence of $\frac{m}{z} = 58$, with no contribution of $\frac{m}{z} =$ 59 and 60, further suggest the presence of TMT. The loss of a -CH$_2$-NH- fragment of TMT will give rise to $\frac{m}{z} = 58$, but no $\frac{m}{z}$ = 59 or 60 fragments can be easily obtained from fragmentation of TMT molecules.\\

The presence of CH$_2$ and NH groups at room temperature and the absence of CH$_3$ (between 3000 - 2948 cm$^{-1}$) and NH$_2$ (only one band is observed between 3450 - 3300 cm$^{-1}$, while NH$_2$ groups would provide the symmetric and antisymmetric N-H stretching modes) absorption features at this temperature, as shown in Fig. \ref{Fig.IR_TPD_residuo}, suggests that \textbf{Dimer A} reacts to form larger species. The addition of an aminomethyl, NH$_2$CH$_3$, group produces TMT. The latter was observed to co-desorb with \textbf{Dimer C} at 212 K, but also at 290 K, by its molecular mass fragment, $\frac{m}{z} = 87, 86$ and $85$ (see Fig. \ref{Fig.QMS_TMT_dimers}).\\

To further confirm the formation of the dimers and larger species upon UV irradiation, the warming up of a non-irradiated methylamine ice is shown in Fig. \ref{Fig.QMS_TMT_blanco}. The ion current measured for $\frac{m}{z} = 58-62$ is different to the one recorded for irradiated methylamine. Covalent dimers have a $\frac{m}{z} = 60$, therefore, their fragmentation give rise to $\frac{m}{z}$ ratios below or equal to 60. From Fig. \ref{Fig.QMS_TMT_blanco}, it can be concluded that $\frac{m}{z} = 61$ and $62$ were present during the desorption with negligible contribution from $\frac{m}{z}$ = 58, 59, and 60. The most likely explanation is that the absence of radical species prevents the formation of covalent dimers, but hydrogen bond interaction will hold pairs of molecules together, forming non-covalent (CH$_3$NH$_2$)$_2$ dimers.\\

\begin{figure}
  \centering 
  \includegraphics[width=0.46\textwidth]{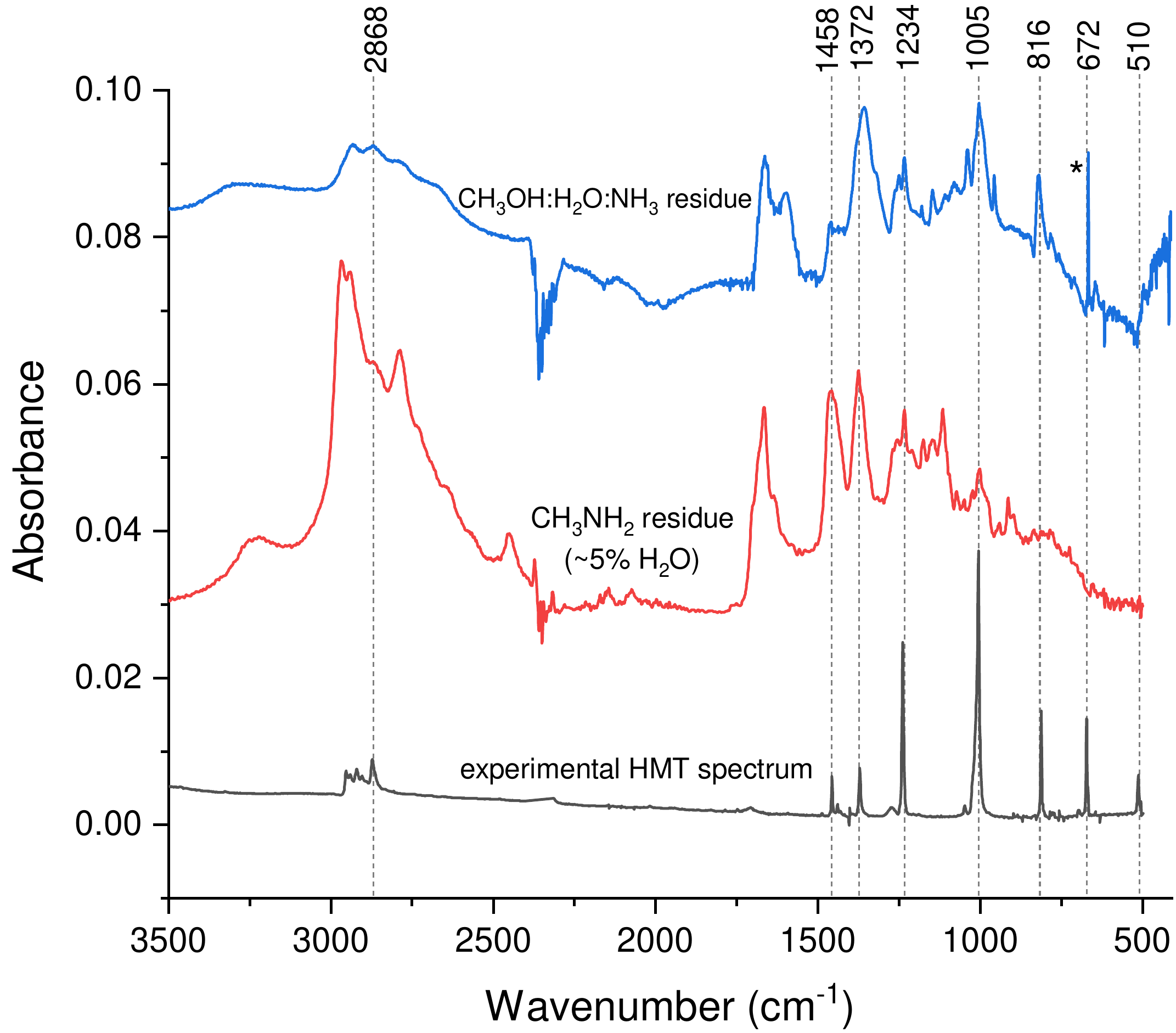}
  \caption{Bottom: Experimental IR spectrum of HMT shown for comparison. Middle: IR spectrum of an irradiated methylamine ice warmed up at 300 K (Exp. \textbf{8}). Top: IR spectrum of an irradiated H$_2$O:CH$_3$OH:NH$_3$ (20:1:1) ice sample warmed up at 300 K. The band labelled with an * is due to impurities present in the substrate.}
  \label{Fig.HMT_formation_comparison}
\end{figure} %

\begin{figure}
  \centering 
  \includegraphics[width=0.4\textwidth]{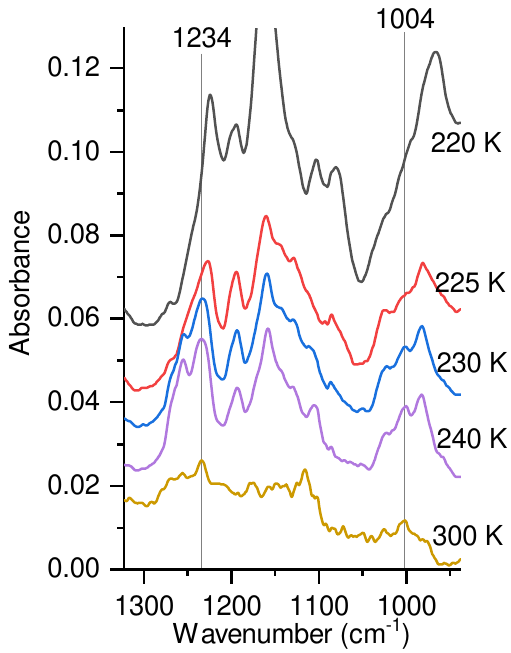}
  \caption{IR spectra during the warm-up of an irradiated CH$_3$NH$_2$ ice. IR features related to HMT at 1004 and 1234 cm$^{-1}$ become clearly visible from 230 K.}
  \label{Fig.formation_HMT_230K}
\end{figure}

Possible formation of HMT in CH$_3$NH$_2$ ice samples was also explored. Fig. \ref{Fig.HMT_formation_comparison} shows the IR spectrum of a CH$_3$NH$_2$ ice sample at 300 K. Comparison with the experimental HMT spectrum suggests that HMT is formed in low H$_2$O containing CH$_3$NH$_2$ ice samples. The organic residue remaining from the H$_2$O:CH$_3$OH:NH$_3$ ice sample, shown in Fig. \ref{Fig.HMT_formation_comparison}, displays more intense bands which can be related to HMT (816, 1005, 1234, 1458, and 2868 cm$^{-1}$), reinforcing the formation of HMT through the mechanism show in Fig. \ref{Fig.sintesis_HMT}. As shown in Fig. \ref{Fig.formation_HMT_230K}, 1004 and 1234 cm$^{-1}$ IR features can be detected from 230 K, confirming that TMT can react to form HMT at this relatively low temperature, what may explain the absence of IR bands related to TMT, as compared to the theoretical spectrum in Fig. \ref{Fig.simulaciones}. Increasing temperatures, as explained above, cause the thermal desorption of remaining TMT molecules, while HMT remains in the refractory residue.\\

No HMT formation was observed in CH$_3$NH$_2$:H$_2$O ice samples. The presence of H$_2$O molecules drives the chemistry of methylamine towards the formation of oxygenated compounds. Water molecules inhibit polymerization of methylenimine, CH$_2$NH, see Fig. \ref{Fig.hidrolisis_imina}. Instead, CH$_2$NH is hydrolyzed to form formamide and formaldehyde. CH$_3$OH:H$_2$O:NH$_3$ ice samples, however, showed an increased HMT formation upon addition of water. H$_2$CO is very efficiently formed from CH$_3$OH molecules, and H$_2$O is required to avoid formaldehyde polymerization. Thus, the presence of H$_2$O favours, in this case, condensation reaction between H$_2$CO and NH$_3$ to produce CH$_2$NH, which cannot be obtained by other routes.\\

\section{Conclusions}
\label{sect.conclusions}
The reactivity of CH$_3$NH$_2$ ice under interstellar conditions was studied in the pure form, in an inert matrix (Xe), and mixed with H$_2$O in the ice, to mimic a more realistic astrophysical scenario. Methylamine-bearing ices were grown at low temperature (8 K) under UHV, and submitted to UV irradiation. From this work, several conclusions can be extracted:

\begin{itemize}
    \item Methylamine thermal desorption is strongly dependent on its environment in the ice. The intermolecular forces established between methylamine and other species may cause thermal desorption to shift from 106 K to more than 150 K.\\
    \item The higher destruction rate of methylamine ice in an inert matrix suggests that a larger fraction of radicals can recombine in pure CH$_3$NH$_2$ ice to reform the parent methylamine molecule. Xe isolated CH$_3$NH$_2$, however, does not trap hydrogen atoms, preventing reformation of CH$_3$NH$_2$ molecules.\\
    \item H$_2$ molecules were not found to desorb when hydrogen bonds are established. Therefore, CH$_3$NH$_2$ and CH$_3$NH$_2$:H$_2$O ice samples showed lower H$_2$ subtraction compared to isolated CH$_3$NH$_2$ molecules in the CH$_3$NH$_2$:Xe ice mixture. Irradiation of pure CH$_4$ ice samples, which cannot form hydrogen bonds, showed an ion current for H$_2$ molecules 100 times higher than the one measured in irradiated CH$_3$NH$_2$ pure ice.\\
    \item The retention of hydrogen within the ice samples leads to the formation of hydrogenated compounds in CH$_3$NH$_2$ and CH$_3$NH$_2$:H$_2$O ice samples over matrix isolated CH$_3$NH$_2$:Xe ice sample.\\
    \item As suggested by, e. g., \cite{MD2018} and \cite{Carrascosa2020MNRAS.493..821C}, CH$\cdot$ radicals are more reactive at low temperatures than NH$\cdot$ radicals. As a result, CH$_3$- bearing compounds are abundant in the irradiated ice samples. CH$_3$CHNH, CH$_3$CN, CH$_3$CH$_2$NH$_2$ were detected in irradiated CH$_3$NH$_2$ ice samples.\\
    \item Oxygenated compounds usually exhibit larger IR band strengths, facilitating their detection. OCN$^{-}$, H$_2$CO and HCONH$_2$ were also detected in the IR spectra of the CH$_3$NH$_2$:H$_2$O ice mixture.\\
    \item At increasingly higher temperatures, the higher mobility of radicals enhances the formation of larger species. The formation of the three possible methylamine dimers (CH$_3$NHNHCH$_3$, NH$_2$CH$_2$CH$_2$NH$_2$ and CH$_3$NHCH$_2$NH$_2$) was discussed. NH$_2$CH$_2$CH$_2$NH$_2$ was found to be the most favoured one, followed by CH$_3$NHCH$_2$NH$_2$. TPD experiments showed the formation of the three dimers, desorbing around 140 K, 210 K and 244 K.\\
    \item Furthermore, comparing irradiated and non-irradiated methylamine ices, it was found that covalent dimers are obtained from UV-irradiation, and non-covalent (CH$_3$NH$_2$)$_2$ dimers are present even without any irradiation period.\\
    \item TMT was detected by thermal codesorption with NH$_2$CH$_2$CH$_2$NH$_2$ and pure thermal desorption at 290 K. The presence of TMT and the relatively low abundance of CH$_3$NHCH$_2$NH$_2$ suggests that TMT may be formed from the reaction between this dimer and an aminomethyl ($\cdot$CH$_2$NH$_2$) species.\\
    \item The absence of IR features that could be related to TMT suggest the rapid reaction of TMT molecules. Indeed, the formation of HMT was detected in UV irradiated CH$_3$NH$_2$-bearing ices at 230 K, long before HMT formation in the 'classical' H$_2$O:CH$_3$OH:NH$_3$ ice mixture, where HMT synthesis takes place near room temperature \citep{Guille2003AA...412..121M, Vinogradoff2012}.\\
    \item The addition of H$_2$O molecules to CH$_3$NH$_2$ inhibits HMT formation. Water molecules react with methylenimine, driving the chemistry towards oxygenated compounds, such as formamide and formaldehyde. On the contrary, CH$_3$OH:H$_2$O:NH$_3$ ice mixtures require the presence of water molecules to prevent formaldehyde polimerization, enhancing the formation of HMT. Therefore, depending on the ice composition, water molecules play a different role regarding HMT formation.\\
    
\end{itemize}

\section{Astrophysical implications}
\label{sect.astrophysical_implications}
The UV absorption cross section of CH$_3$NH$_2$ ice within the 120-180 nm range was reported in this work (see Fig. \ref{Fig.VUV_comparison}), measuring a value of 5.6$\pm$0.8$\times$10$^{-18}$ molecule cm$^{-2}$ for pure CH$_3$NH$_2$ ice. The UV absorption spectrum makes it possible to calculate the photon energy absorbed by methylamine ice which can be incorporated into computational models to improve the accuracy of the simulations.\\

CH$_3$NH$_2$ is of particular interest in astrochemistry, as it contains a C-N bond, which is present in most biological molecules such as aminoacids, DNA, or heterocyclic compounds. The first detection of methylamine in the gas phase of the interstellar medium by \cite{Kaifu1974ApJ...191L.135K} opened the possibility to include methylamine in solid phase ice models. Experimental simulations including methylamine have shown its potential to give rise to COMs \citep{Holtom2005ApJ...626..940H, Bossa2009AA...506..601B, Lee2009ApJ...697..428L, Vinogradoff2013}. \cite{Vinogradoff2012} and references therein studied the formation mechanism of HMT and complex mixtures containing methylamine. This work focuses on the reactivity of pure methylamine ice at low temperature exposed to UV irradiation, as a primary step to understand its chemistry toward the formation of N-heterocycles, in particular TMT and HMT. The addition of water molecules approaches to a more realistic astrophysical scenario of ice mantle photoprocessing followed by warm-up.\\

Several COMs were formed (see Table \ref{Table.fotoproductos_irr}). Among them, formamide, HCONH$_2$, is one of the most studied ones due to its prebiotic potencial to form a full variety of species which are known to be present in our metabolism \citep{SaladinoC2CS35066A}. The addition of H$_2$O to the ice sample enhances the formation of HCONH$_2$, which was first detected in the ISM in Sgr B2 molecular cloud \citep{Rubin1971ApJ...169L..39R}, and later on in 67P/Churyumov-Gerasimenko comet during the Rosetta mission \citep{Goesmann2015Sci...349b0689G, Altwegg2017MNRAS.469S.130A}. H$_2$O molecules, as it can be derived from Fig. \ref{Fig.destruccion_metilamina}, lead to a large overall dissociation of methylamine molecules, producing a more complex chemistry.\\

\cite{Vinogradoff2012} reported the formation mechanism of HMT starting from a ternary H$_2$CO:NH$_3$:HCOOH ice mixture. They found evidence for the formation of the protonated form of TMT (TMTH$^{+}$) as an intermediate species during HMT synthesis. They stated that HCOOH is crucial to stabilize CH$_2$NH allowing its presence at larger temperatures, where TMT can be thermally formed. The detection of TMT in pure CH$_3$NH$_2$ experiments, and the absence of HMT formation suggest that a complex synthesis is required to form HMT, in agreement with previous works \citep[][and references therein]{Vinogradoff2012}. Furthermore, this study deeps in the formation of TMT, which can lead to the synthesis of other N-heterocycles under astrophysically relevant scenarios, similar to those observed from H$_2$O:CH$_3$OH/CO:NH$_3$ ice mixtures \cite[see, for example][]{Meierhenrich2005_N-heterocycles, Oba2019NatCo..10.4413O}. Formaldehyde, H$_2$CO, which is negligible in our experiments, but efficiently formed from methanol, is required for the formation of more complex heterocycles \citep{Guille2003AA...412..121M, Vinogradoff2012}.\\

Even though methylamine is a minor component in the gas of the interstellar medium, after accretion on ice mantles or direct formation in the ice, thermal desorption of other species in hot cores or warm protoplanetary disk regions can increase the relative abundance of methylamine in ice mantles. Indeed, prior to the thermal desorption of methylamine molecules, other species, such as CO, CH$_4$, NH$_3$, O$_2$, N$_2$ and CO$_2$, should have sublimated to the gas phase. Therefore, UV-irradiation and thermal processing of the ice mantles can increase the relative abundance of methylamine photoproducts, promoting chemical reactions between them which may lead to the formation of COMs following the synthetic pathways presented in this work.\\

\section{Acknowledgements}
We would like to devote this paper to the memory of our long-term colleague and dearest friend professor Rafael Escribano from Spanish CSIC, whose advise and stimulating discussions we enjoyed so much. The Spanish Ministry of Science, Innovation and Universities supported this research under grant number AYA2017-85322-R and MDM-2017-0737 Unidad de Excelencia "Mar\'{i}a de Maeztu"-- Centro de Astrobiolog\'{i}a (CSIC-INTA). (AEI/FEDER, UE). P. C. G. acknowledges support from PGC2018-096444-B-I00. H. C. was supported by PhD fellowship FPU-17/03172. M. L S. acknowledges Ministerio de Econom\'ia y Competitividad of Spain for the project AGL2016-80475-R and Comunidad Aut\'onoma of Madrid (Spain) and European funding from FEDER program (project S2018/BAA-4393, AVANSECAL-II-CM).\\

\section{Data availability}
The data underlying this article will be shared on reasonable request to the corresponding authors.\\

\bibliographystyle{mnras}
\bibliography{bibliography} 

\bsp	
\label{lastpage}







\end{document}